\documentclass[aip, jcp, preprint,showpacs,superscriptaddress,groupedaddress]{revtex4-1}

\usepackage{amsmath}
\usepackage{amssymb}
\usepackage{braket}
\usepackage{bm}

\usepackage{multirow,bigdelim}

\usepackage{graphicx}
\usepackage{subcaption}

\newtheorem{theorem}{Theorem}[section]
\newtheorem{lemma}[theorem]{Lemma}

\begin{document}

\title{Reduced Density Matrices / Static Correlation Functions of Richardson-Gaudin States Without Rapidities}
\author{Alexandre Faribault}
 \email{alexandre.faribault@univ-lorraine.fr}
\author{Claude Dimo}
\affiliation{Universit\'{e} de Lorraine, CNRS, LPCT, F-54000 Nancy, France}

\author{Jean-David Moisset and Paul A. Johnson}
 \email{paul.johnson@chm.ulaval.ca}
 \affiliation{D\'{e}partement de chimie, Universit\'{e} Laval, Qu\'{e}bec, Qu\'{e}bec, G1V 0A6, Canada}

\date{\today}

\begin{abstract}
Seniority-zero geminal wavefunctions are known to capture bond-breaking correlation. Among this class of wavefunctions, Richardson-Gaudin states stand out as they are eigenvectors of a model Hamiltonian. This provides a clear physical picture, clean expressions for reduced density matrix (RDM) elements, and systematic improvement (with a complete set of eigenvectors). Known expressions for the RDM elements require the computation of rapidities, which are obtained by first solving for the so-called eigenvalue based variables (EBV) then root-finding of a Lagrange interpolation polynomial. In this manuscript we obtain expressions for the RDM elements directly in terms of the EBV. The final expressions can be computed with the same cost as the rapidity expressions. Therefore, except in particular circumstances, it is entirely unnecessary to compute rapidities at all. The RDM elements require numerically inverting a matrix and while this is usually undesirable we demonstrate that it is stable, except when there is degeneracy in the single-particle energies. In such cases a different construction would be required.
\end{abstract}

\maketitle

\section{Introduction}
The majority of systems in quantum chemistry are weakly-correlated: the electrons form a mean-field around the potential dominated by the nuclei. The wavefunction is well-described by a short expansion of Slater determinants from a single reference. Kohn-Sham density functional theory (DFT) and coupled cluster (CC) with singles and doubles usually provide quantitatively correct results.\cite{helgaker_book}

Strongly-correlated systems are broadly defined as those that are not weakly-correlated: the wavefunction cannot be built from a single reference Slater determinant. If it is understood which Slater determinants are the most important, then the complete active space self-consistent field (CASSCF) or complete active space configuration interaction (CASCI) are good approaches, but become less effective when the number of important Slater determinants becomes large. State of the art algorithms\cite{white:1992,white:1993,chan:2002,chan:2004,chan:2011,thom:2005,booth:2010,booth:2013,huron:1973,sharma:2017,holmes:2017,li:2018,yao:2021} are able to treat larger strongly-correlated systems, but are still fixed in the picture of weakly interacting electrons.

Weakly interacting \emph{pairs} of electrons, geminals, are known to provide a better starting point for many strongly-correlated systems.\cite{fock:1950,mcweeny:1959,mcweeny:1960,mcweeny:1963,nicely:1971,siems:1976} For systems with no unpaired electrons, a general closed-shell pair mean-field, the antisymmetrized product of interacting geminals (APIG),\cite{silver:1969,silver:1970a,silver:1970b,silver:1970c} is near-exact\cite{moisset:2022a} but is not feasible variationally nor by projection though a few of its special cases are. The antisymmetrized geminal power (AGP)\cite{coleman:1965,ortiz:1981,sarma:1989,coleman:1997,henderson:2019,khamoshi:2019,dutta:2020,khamoshi:2021,dutta:2021} is variationally feasible, though unless Jastrow factors are included AGP is not size-consistent.\cite{neuscamman:2012,neuscamman:2013,neuscamman:2016} This limits AGP's utility for molecular systems. The antisymmetrized product of strongly-orthogonal geminals (APSG)\cite{hurley:1953,kutzelnigg:1964} and the generalized valence bond/perfect pairing (GVB)\cite{goddard:1967,hay:1972,hunt:1972,goddard:1973} are variationally feasible and treat the dissociated limit of molecules correctly.\cite{kutzelnigg:2010,kobayashi:2010,kutzelnigg:2012,surjan:2012,zoboki:2013,pernal:2014,jeszenszki:2014,pastorczak:2015,margocsy:2018,pernal:2018,pastorczak:2018,pastorczak:2019,piris:2011} APSG and GVB require assigning orbitals into disjoint subspaces which is in general difficult. The antisymmetrized product of 1-reference orbital geminals (AP1roG),\cite{limacher:2013} equivalent to pair-coupled-cluster doubles (pCCD),\cite{stein:2014} can be solved by projection with $\mathcal{O}(N^4)$ scaling, with $N$ the number of spatial orbitals. AP1roG/pCCD has shown quite promising results for repulsive Coulomb systems, molecular dissociations in particular.\cite{limacher:2014a,limacher:2014b,henderson:2014a,henderson:2014b,boguslawski:2014a,boguslawski:2014b,boguslawski:2014c,tecmer:2014,boguslawski:2015,marie:2021,kossoski:2021} It is however a state specific method that must be solved by projection.

Recently, we have employed the eigenvectors of the reduced Bardeen-Cooper-Schrieffer (BCS)\cite{bardeen:1957a,bardeen:1957b,schrieffer_book} Hamiltonian, the so-called Richardson-Gaudin (RG) states, as a variational wavefunction ansatz for molecular dissociations. RG states are a particular case of APIG for which the geminal coefficients are parametrized by \emph{rapidities} that solve non-linear equations. As they are eigenvectors of a model system, their 1- and 2-body reduced density matrix (RDM) elements, or in the condensed-matter literature their \emph{static correlation functions}, are computable with a reasonable cost. However, what sets RG states apart from other degenerate cases of APIG is that they form a basis for the Hilbert space. Therefore, even if a single RG state is insufficient, systematic improvement is achieved by adding more states to the expansion of the wavefunction.

Employing a wavefunction ansatz variationally requires practical formulas for its RDM elements. Expressions for the 1- and 2-RDM elements require numerically computing rapidities. This entails solving a set of non-linear equations for the so-called eigenvalue based variables (EBV) then locating the roots of a Lagrange interpolation polynomial.\cite{faribault:2011,elaraby:2012} In this contribution, we demonstrate that the RDM elements are computable directly from the EBV. This approach is more stable numerically, avoids many unnecessary computations and eliminates a source of numerical instability.

Section \ref{sec:rg_states} summarizes the relevant properties of RG states and presents their RDM elements in terms of rapidities. In section \ref{sec:ebv} we obtain expressions for the RDM elements in terms of the EBV, along with their derivatives. As we are presenting the general case, keeping track of the signs is an incredibly tedious task. We encourage the interested reader to perform the calculation for a specific element which is much more clear. Our formulas require numerically inverting a matrix, which is a task that is typically avoided. As such, in section \ref{sec:stability} we demonstrate that the condition number of the matrix is small enough that the numerical inverse is reasonable unless the single-particle energies are degenerate.

\section{RG states} \label{sec:rg_states}
\subsection{Reduced BCS Hamiltonian}
Pairs of electrons are built with the Lie algebra su(2). In particular, the three objects
\begin{align} \label{eq:su2_objects}
S^+_i = a^{\dagger}_{i\uparrow} a^{\dagger}_{i\downarrow}, \quad S^-_i = a_{i\downarrow} a_{i\uparrow}, \quad
S^z_i = \frac{1}{2} \left( a^{\dagger}_{i\uparrow}a_{i\uparrow} + a^{\dagger}_{i\downarrow}a_{i\downarrow} -1 \right)
\end{align}
create ($S^+_i$), remove ($S^-_i$), and count the number of pairs ($S^z_i$) in the spatial orbital $i$. The second quantized operators $a^{\dagger}_{i\uparrow}$ create an up-spin electron in spatial orbital $i$ etc. The su(2) operators have the structure 
\begin{subequations} \label{eq:su2_structure}
\begin{align} 
[S^+_i , S^-_j] &= 2 \delta_{ij} S^z_i \\
[S^z_i , S^{\pm}_j] &= \pm \delta_{ij} S^{\pm}_i.
\end{align}
\end{subequations}
It is also convenient to use the number operator
\begin{align}
\hat{n}_i = 2S^z_i + 1.
\end{align}
Richardson\cite{richardson:1963,richardson:1964,richardson:1965} and Gaudin\cite{gaudin:1976} showed that the reduced BCS Hamiltonian
\begin{align} \label{eq:bcs_ham}
\hat{H}_{BCS} = \frac{1}{2} \sum^N_{i=1} \varepsilon_i \hat{n}_i - \frac{g}{2} \sum_{ij} S^+_i S^-_j
\end{align}
has a complete set of structured eigenvectors
\begin{align} \label{eq:rg_states}
\ket{\{u\}} = S^+(u_1) S^+(u_2) \dots S^+(u_M) \ket{\theta}.
\end{align}
We will refer to the states \eqref{eq:rg_states} as RG states. The $M$ operators $S^+(u_a)$ create pairs de-localized over the set of $N$ available spatial orbitals
\begin{align}
S^+(u) = \sum^N_{i=1} \frac{S^+_i}{u - \varepsilon_i}.
\end{align}
The vacuum $\ket{\theta}$ is chosen so that it is destroyed by each of the pair-removal objects
\begin{align}
S^-_i \ket{\theta} = 0, \quad \forall i.
\end{align}
Usually we consider $\ket{\theta}$ to be the physical vacuum, but it may be chosen more generally to include any state that does not participate in the pairing structure in \eqref{eq:su2_structure}.  The RG states \eqref{eq:rg_states} are eigenvectors of \eqref{eq:bcs_ham} provided that the set of complex numbers $\{u\}$, ordinarily called the \emph{rapidities}, are solutions of the set coupled non-linear equations 
\begin{align} \label{eq:rich_eq}
\frac{2}{g} + \sum^N_{i=1} \frac{1}{u_a - \varepsilon_i} + \sum_{b (\neq a)} \frac{2}{u_b - u_a} = 0, \quad \forall a = 1,\dots,M
\end{align}
which are known as Richardson's equations. Many algorithms exist,\cite{rombouts:2004,guan:2012,pogosov:2012,debaerdemacker:2012,claeys:2015} though solving directly for the rapidities is not the best approach as these equations have divergent critical points where one of the rapidities coincides with one of single-particle energies $\{\varepsilon\}$. A far more stable approach is to define the EBV
\begin{align}
U_i = \sum_a \frac{1}{\varepsilon_i - u_a},
\end{align}
and notice that Richardson's equations are equivalent to the set of equations for the EBV
\begin{align} \label{eq:ebv_eq}
U^2_i - \frac{2}{g} U_i - \sum_{k \neq i} \frac{U_k - U_i}{\varepsilon_k - \varepsilon_i} = 0, \quad \forall i = 1,\dots,N,
\end{align}
from which the rapidities may be obtained with a root-finding procedure based on Lagrange interpolation.\cite{faribault:2011,elaraby:2012} In our variational calculations\cite{fecteau:2022} we have found this approach to be by far the most reliable. While solving the non-linear equations \eqref{eq:ebv_eq}, it is important to enforce the normalization
\begin{align}
\sum_i U_i = \frac{2M}{g}
\end{align}
as otherwise the solutions will cross into different particle number sectors. Since the equations \eqref{eq:ebv_eq} no longer have variables in the denominators, they are much easier to solve numerically without worrying about critical points. To compute rapidities we must first compute the EBV, so expressions for the RDM elements directly in terms of the EBV would eliminate the need to compute the rapidities at all.

\subsection{Coulomb Energy Functional}
We are using RG states as variational trial functions, in particular for Coulomb Hamiltonians
\begin{align} \label{eq:C_ham}
\hat{H}_C = \sum_{ij} h_{ij} \sum_{\sigma} a^{\dagger}_{i \sigma} a_{j \sigma} + \frac{1}{2} \sum_{ijkl} V_{ijkl} \sum_{\sigma \tau} a^{\dagger}_{i \sigma} a^{\dagger}_{j \tau} a_{l \tau} a_{k \sigma}
\end{align}
describing molecular systems. The one- and two-electron integrals
\begin{align}
h_{ij} &= \int d\mathbf{r} \phi^*_i (\mathbf{r}) \left( - \frac{1}{2} \nabla^2 - \sum_I \frac{Z_I}{| \mathbf{r} - \mathbf{R}_I |} \right) \phi_j (\mathbf{r}) \\
V_{ijkl} &= \int d\mathbf{r}_1 d\mathbf{r}_2 \frac{\phi^*_i(\mathbf{r}_1)  \phi^*_j(\mathbf{r}_2)  \phi_k(\mathbf{r}_1)  \phi_l(\mathbf{r}_2)  }{| \mathbf{r}_1 - \mathbf{r}_2|}
\end{align}
are expressed in a basis of functions $\{\phi\}$. Evaluating the expected value of the energy of \eqref{eq:C_ham} with an RG state gives
\begin{align} \label{eq:e_functional}
E[\{\varepsilon\},g] = 2\sum_k h_{kk} \gamma_k + \sum_{kl} (2V_{klkl}-V_{kllk})D_{kl} + \sum_{kl} V_{kkll} P_{kl},
\end{align}
where the 1-RDM
\begin{align}
\gamma_k = \frac{1}{2} \frac{\braket{\{u\}|\hat{n}_k|\{u\}}}{\braket{\{u\}|\{u\}}}
\end{align}
is diagonal, and the only non-zero elements of the 2-RDM are the \emph{diagonal-correlation function}
\begin{align}
D_{kl} = \frac{1}{4} \frac{\braket{\{u\}|\hat{n}_k \hat{n}_l|\{u\}}}{\braket{\{u\}|\{u\}}}
\end{align}
and the \emph{pair-correlation function}
\begin{align}
P_{kl} = \frac{\braket{\{u\}|S^+_k S^-_l | \{u\}}}{\braket{\{u\}|\{u\}}}.
\end{align}
Note that the diagonal elements $D_{kk}$ and $P_{kk}$ refer to the same element ($\gamma_k$), and so by convention we assign it to $P_{kk}=\gamma_k$ and set $D_{kk}=0$. The energy \eqref{eq:e_functional} is a functional of the parameters $\{\varepsilon\}$ and $g$ which define a reduced BCS Hamiltonian, along with the particular choice of RG state. 

\subsection{Rapidity based scalar products}
Scalar products and density matrix elements in terms of rapidities have been computed many times.\cite{amico:2002,faribault:2008,faribault:2010,gorohovsky:2011,fecteau:2020} Hence, we will \emph{very} briefly summarize the results. We begin with two RG states \eqref{eq:rg_states}: one \emph{on-shell} with rapidities $\{u\}$ a solution of Richardson's equations \eqref{eq:rich_eq} and a second \emph{off-shell} with arbitrary rapidities $\{v\}$. The pertinent form of Slavnov's theorem\cite{slavnov:1989,zhou:2002} gives the scalar product as a single determinant
\begin{align} \label{eq:slavnov_rap}
\braket{\{u\} | \{v\}} = P(\{u\},\{v\}) \det L(\{u\},\{v\})
\end{align}
where $P$ 
\begin{align}
P (\{u\},\{v\}) = \frac{\prod_{ab} (u_a - v_b)}{\prod_{a<b} (v_a - v_b)(u_b - u_a)}
\end{align}
is the reciprocal of the determinant of a Cauchy matrix of $\{u\}$ and $\{v\}$, while the matrix $L$ is
\begin{align}
L_{ab} = \frac{1}{(u_a - v_b)^2} \left( \frac{2}{g} + \sum_i \frac{1}{v_b - \varepsilon_i} - \sum_{c \neq a} \frac{2}{(v_b - u_c)} \right). 
\end{align}
When $\{u\}=\{v\}$ this reduces to
\begin{align}
\braket{\{u\}|\{u\}} = \det G
\end{align}
with $G$ the \emph{Gaudin} matrix
\begin{align} \label{eq:Gmat}
G_{ab} = 
\begin{cases}
\sum_i \frac{1}{(u_a - \varepsilon_i)^2} - \sum_{c(\neq a)} \frac{2}{(u_a - u_c)^2}, \quad & a = b\\
\frac{2}{(u_a-u_b)^2}, & a \neq b
\end{cases}
\end{align}
which is the Jacobian of Richardson's equations \eqref{eq:rich_eq}.

RDM elements, are evaluated with the form factor approach: to calculate the 1-RDM elements, the commutator
\begin{align}
[\hat{n}_k, S^+(v)] = \frac{2S^+_k}{v-\varepsilon_k}
\end{align}
is used to move $\hat{n}_k$ to the right, past each $S^+(v_a)$, until it destroys the vacuum, giving a sum of scalar products 
\begin{align} \label{eq:1rdm_ff}
\frac{1}{2} \braket{\{u\} | \hat{n}_k | \{v\}} = \sum_a \frac{\braket{\{u\} |S^+_k | \{v\}_a }}{(v_a - \varepsilon_k)}
\end{align}
that may be evaluated as limits of Slavnov's theorem. In equation \eqref{eq:1rdm_ff}, the notation $\{v\}_a$ means the set $\{v\}$ without the element $v_a$. For the sake of simplicity we will refer to the scalar product in the numerator as a form factor, though usually that would imply that $\{v\}$ were also a solution of Richardson's equations. Form factors are evaluated as limits of Slavnov's theorem. Notice that the local pair creators are the residues of the RG pair creators at each of the simple poles
\begin{align}
S^+_k = \lim_{v \rightarrow \varepsilon_k} (v - \varepsilon_k) S^+(v),
\end{align} 
so that the form factor is the residue of Slavnov's theorem
\begin{align} \label{eq:ff1}
\braket{\{u\} | S^+_k | \{v\}_a} = \lim_{v_a \rightarrow \varepsilon_k} (v_a - \varepsilon_k) \braket{\{u\} | \{v\}}.
\end{align}
Taking the residue, and setting $\{v\}=\{u\}$ gives
\begin{align}
\braket{\{u\}|S^+_k | \{u\}_a } = (u_a-\varepsilon_k) \det G(a\rightarrow \textbf{b}_k)
\end{align}
where $G(a\rightarrow \textbf{b}_k)$ is the Gaudin matrix \eqref{eq:Gmat} whose $a$th column has been replaced by the vector
\begin{align} \label{eq:rap_rhs}
\textbf{b}_k = \begin{pmatrix}
\frac{1}{(u_1 - \varepsilon_k)^2} \\
\frac{1}{(u_2 - \varepsilon_k)^2} \\
\vdots \\
\frac{1}{(u_M - \varepsilon_k)^2}
\end{pmatrix}.
\end{align}
It is not difficult to see that the derivatives of the rapidities $\{u\}$ with respect to the single-particle energies $\{\varepsilon\}$ are the solutions of the linear equations
\begin{align} \label{eq:rap_lin_eq}
G \frac{\partial \textbf{u}}{\partial \varepsilon_k} = \textbf{b}_k.
\end{align}
Normalizing the form factors amounts to dividing by $\det G$, and the resulting ratios are obtained directly from Cramer's rule as
\begin{align}
\frac{\det G(a\rightarrow \textbf{b}_k)}{\det G} = \frac{\partial u_a}{\partial \varepsilon_k}
\end{align}
giving the final expression for the 1-RDM elements
\begin{align} \label{eq:1dm_rap}
\gamma_k = \sum_a \frac{\partial u_a}{\partial \varepsilon_k}.
\end{align}

As they will be important for the next section, the results for the 2-body density matrix elements in terms of form factors are
\begin{align}
\frac{1}{4} \braket{\{u\} | \hat{n}_k \hat{n}_l | \{v\}} &= \sum_{a\neq b} \frac{\braket{\{u\} | S^+_k S^+_l | \{v\}_{a,b}}}
{(v_a - \varepsilon_k)(v_b - \varepsilon_l)} \label{eq:d2d_ff} \\
\braket{\{u\} | S^+_k S^-_l | \{v\}} &= \sum_a \frac{\braket{\{u\} |S^+_k | \{v\}_a }}{(v_a - \varepsilon_l)}
- \sum_{a \neq b} \frac{\braket{\{u\} | S^+_k S^+_l | \{v\}_{a,b}}}{(v_a - \varepsilon_l)(v_b - \varepsilon_l)} \label{eq:d2p_ff}
\end{align}
where 
\begin{align}
\braket{\{u\} | S^+_k S^+_l | \{v\}_{a,b}} = \lim_{v_a \rightarrow \varepsilon_k}\lim_{v_b \rightarrow \varepsilon_l} (v_a - \varepsilon_k)(v_b - \varepsilon_l) \braket{\{u\}|\{v\}}.
\end{align}
The form factors become
\begin{align}
\braket{\{u\}|S^+_kS^+_l| \{u\}_{a,b}} = \frac{(u_a-\varepsilon_k)(u_b-\varepsilon_k)(u_a-\varepsilon_l)(u_b-\varepsilon_l)}{(\varepsilon_k - \varepsilon_l)(u_b-u_a)} \det G(a,b \rightarrow \textbf{b}_k,\textbf{b}_l)
\end{align}
where $G(a,b \rightarrow \textbf{b}_k,\textbf{b}_l)$ is the Gaudin matrix \eqref{eq:Gmat} with the $a$th column replaced with the $k$th version of \eqref{eq:rap_rhs} and the $b$th column replaced with the $l$th version of \eqref{eq:rap_rhs}. A fundamental result of scaled determinants is that the ratio of two determinants differening by $k$ columns is identical to a $k\times k$ determinant of ratios of two determinants differing by single columns. In the present case this means
\begin{align}
\frac{\det G(a,b \rightarrow \textbf{b}_k,\textbf{b}_l)}{\det G} = \frac{\partial u_a}{\partial \varepsilon_k} \frac{\partial u_b}{\partial \varepsilon_l} - \frac{\partial u_a}{\partial \varepsilon_l} \frac{\partial u_b}{\partial \varepsilon_k}.
\end{align}
Finally,
\begin{align} \label{eq:d2d_rap}
D_{kl} = \sum_{a < b} \frac{(u_a-\varepsilon_k)(u_b-\varepsilon_l) + (u_a-\varepsilon_l)(u_b-\varepsilon_k)}{(\varepsilon_k - \varepsilon_l)(u_b-u_a)} \left( \frac{\partial u_a}{\partial \varepsilon_k} \frac{\partial u_b}{\partial \varepsilon_l} - \frac{\partial u_a}{\partial \varepsilon_l} \frac{\partial u_b}{\partial \varepsilon_k} \right)
\end{align}
and
\begin{align} \label{eq:d2p_rap}
P_{kl} = \sum_a \frac{(u_a-\varepsilon_k)}{(u_a-\varepsilon_l)} \frac{\partial u_a}{\partial \varepsilon_k} 
-2 \sum_{a<b} \frac{(u_a-\varepsilon_k)(u_b-\varepsilon_k)}{(\varepsilon_k - \varepsilon_l)(u_b-u_a)} \left( \frac{\partial u_a}{\partial \varepsilon_k} \frac{\partial u_b}{\partial \varepsilon_l} - \frac{\partial u_a}{\partial \varepsilon_l} \frac{\partial u_b}{\partial \varepsilon_k} \right).
\end{align}

The primitive elements $\frac{\partial \textbf{u}}{\partial \varepsilon_k}$ are obtained by solving the linear equations \eqref{eq:rap_lin_eq} and the 2-RDM can be computed with $\mathcal{O}(N^2 M^2)$ cost: there are $\mathcal{O}(N^2)$ elements and each requires computing a sum with $\mathcal{O}(M^2)$ terms.

\section{EBV Scalar Products and Reduced Density Matrices} \label{sec:ebv}
From two sets of rapidities $\{u\}$ and $\{v\}$ we define the corresponding EBV
\begin{align}
U_i &= \sum_a \frac{1}{\varepsilon_i - u_a} \\
V_i &= \sum_a \frac{1}{\varepsilon_i - v_a}
\end{align}
and henceforth we will always consider the set $\{U\}$ to be a solution to the coupled non-linear equations \eqref{eq:ebv_eq} while the set $\{V\}$ are arbitrary. The purpose of the present contribution is to compute the 1- and 2-RDMs of RG states without rapidities. However, the RG states themselves do not have a simple expression directly in terms of EBV, and hence RG states will be continued to be labelled by rapidities. As shown previously,\cite{faribault:2012,claeys:2017b} the scalar product $\braket{\{u\} | \{v\}}$ has a determinant expression in terms of the EBV
\begin{align} \label{eq:pieter}
\braket{\{u\} | \{v\}} = \eta \det J
\end{align}
with the constant
\begin{align}
\eta = (-1)^{N-M} \left(\frac{g}{2} \right)^{N-2M}
\end{align}
and the matrix $J$
\begin{align}
J_{ij} = \begin{cases}
U_i + V_i - \frac{2}{g} + \sum_{k (\neq i)} \frac{1}{\varepsilon_k - \varepsilon_i}, & i = j \\
\frac{1}{\varepsilon_i - \varepsilon_j} & i \neq j
\end{cases}.
\end{align}
The factor $\eta$ differs from ref.\cite{claeys:2017b} since our definition of $g$ is the negative of theirs. Further, the matrix $J$ is the transpose of that reported. This does not change the determinant but makes the present development simpler in terms of columns. 

When the states are the same, i.e. the rapidities $\{u\} = \{v\}$ and hence the EBV $\{U\} = \{V\}$ are the same, we will specify the matrix $\bar{J}$
\begin{align}
\bar{J}_{ij} = \begin{cases}
2 U_i - \frac{2}{g} + \sum_{k (\neq i)} \frac{1}{\varepsilon_k - \varepsilon_i}, & i = j \\
\frac{1}{\varepsilon_i - \varepsilon_j} & i \neq j
\end{cases}
\end{align}
and highlight that it is the Jacobian of the EBV equations \eqref{eq:ebv_eq}, just like the Gaudin matrix $G$ is the Jacobian of Richardson's equations \eqref{eq:rich_eq}.

We will proceed in the same manner as for rapidities. Individual form factors will be computed as residues of the scalar product \eqref{eq:pieter}, but the development is trickier and requires using two practical lemmas (proofs in appendices \ref{sec:lemma_1} and \ref{sec:lemma_2}) many, many times.

\begin{lemma} \label{lem:mat_diag}
For the matrix $J$, the diagonal rank-$N$ update for the determinant of $J$ to $J-d(z)$ with
\begin{align} \label{eq:diag_mat}
d (z) = \begin{pmatrix}
\frac{1}{\varepsilon_1 - z} & 0 & \dots & 0 \\
0 & \frac{1}{\varepsilon_2 - z} & \dots & 0 \\
\vdots & \vdots & \ddots & \vdots \\
0 & 0 & \dots & \frac{1}{\varepsilon_N - z}
\end{pmatrix}
\end{align}
where $z$ is a complex number distinct from the set $\{\varepsilon\}$, is equivalent to the rank-one update
\begin{align}
\det (J - d(z)) = \det (J - \textbf{x}(z) \textbf{1}^T).
\end{align}
The $N$-element vectors being
\begin{align}
\textbf{x}(z)^T &= \begin{pmatrix} \frac{1}{\varepsilon_1 - z} & \frac{1}{\varepsilon_2 - z} & \dots & \frac{1}{\varepsilon_N - z} \end{pmatrix} \\
\textbf{1}^T &= \begin{pmatrix} 1 & 1 & \dots & 1 \end{pmatrix}.
\end{align}
\end{lemma}

\begin{lemma} \label{lem:update}
For an invertible $N\times N$ matrix $J$, the sum of $M$ rank-one updates with arbitrary vectors $\textbf{x}_j$ and a unique vector $\textbf{y}^T$ is
\begin{align}
\sum^M_{j=1} \alpha_j \det (J - \textbf{x}_j\textbf{y}^T) = \det \begin{pmatrix}
\sum^M_{j=1} \alpha_j & \textbf{y}^T \\
\sum^M_{j=1} \alpha_j \textbf{x}_j & J
\end{pmatrix},
\end{align}
with scalars $\alpha_j$.
\end{lemma}
With these two lemmas we can proceed to calculating form factors and density matrix elements in terms of the EBV.

\subsection{1-RDM elements}
The two formulas for the scalar product \eqref{eq:slavnov_rap} and \eqref{eq:pieter} are equivalent, so the scalar product \eqref{eq:pieter} must have the same simple poles and residues as \eqref{eq:slavnov_rap}. The corresponding form factor is once again given by \eqref{eq:ff1}, but the residue \emph{appears} to be different. Only the diagonal element $J_{kk}$ has the simple pole, with residue
\begin{align} \label{eq:residue}
\lim_{v_a \rightarrow \varepsilon_k} (v_a - \varepsilon_k) J_{kk} = -1.
\end{align}
Expansion along the $k$th column (or $k$th row) thus yields only one term, while the other diagonal elements are modified, for $k\neq i$
\begin{align} 
\lim_{v_a \rightarrow \varepsilon_k} J_{ii} &= U_i - \frac{2}{g} + \sum_{b (\neq a)} \frac{1}{\varepsilon_i - v_b} + \frac{1}{\varepsilon_i - \varepsilon_k} + \sum_{j (\neq i)} \frac{1}{\varepsilon_j - \varepsilon_i} \\
&= J_{ii} + \frac{1}{\varepsilon_i - \varepsilon_k} - \frac{1}{\varepsilon_i - v_a}, \label{eq:diag_update}
\end{align}
where in the first line the limit is taken, and in the second line we have added and subtracted $\frac{1}{\varepsilon_i - v_a}$ to write the expression as an update to the matrix element $J_{ii}$. The form factor \eqref{eq:ff1} is thus the determinant of the $(N-1)\times (N-1)$ matrix obtained from $J$ by removing the $k$th row and the $k$th column, denoted $J^{k,k}$, and updating the remaining diagonal elements as in \eqref{eq:diag_update}, giving
\begin{align}
\braket{\{u\} | S^+_k | \{v\}_a } = \eta \det \left( J^{k,k} + d^{k,k} (\varepsilon_k) - d^{k,k} (v_a) \right)
\end{align}
where $d^{k,k}(z)$ is \eqref{eq:diag_mat} without the $k$th row and the $k$th column. Using lemma \ref{lem:mat_diag}, this determinant can be written as a rank-1 update
\begin{align}
\det \left( J^{k,k} + d^{k,k} (\varepsilon_k) - d^{k,k} (v_a) \right) = \det \left( J^{k,k} + d^{k,k} (\varepsilon_k) - \textbf{x}^k (v_a) \textbf{1}^T \right)
\end{align}
with the $N-1$ element vectors $\textbf{x}^k(v_a)^T=\begin{pmatrix}
\frac{1}{\varepsilon_1 - v_a} & \frac{1}{\varepsilon_2 - v_a} & \dots & \frac{1}{\varepsilon_N - v_a}
\end{pmatrix}$ (\emph{without} the $k$th element) and $\textbf{1}^T = \begin{pmatrix}
1 & 1 & \dots 1
\end{pmatrix}$. The sum \eqref{eq:1rdm_ff} can now be evaluated
\begin{align} \label{eq:1rdm_J_sum}
\frac{1}{2}\braket{\{u\} | \hat{n}_k | \{v\}} = \eta \sum_a \frac{\det \left(J^{k,k} +d^{k,k}(\varepsilon_k) -\textbf{x}^k(v_a)\textbf{1}^T \right)}{\varepsilon_k - v_a}
\end{align}
where the negative sign from the residue of the pole \eqref{eq:residue} has been absorbed by switching the denominator. The RHS of \eqref{eq:1rdm_J_sum} is a sum of rank-1 updates of the common $(N-1)\times (N-1)$ matrix $J^{k,k}$ which, using lemma \ref{lem:update}, is equivalent to a single determinant of the rank $N$ matrix
\begin{align}
\sum_a \frac{\det \left(J^{k,k} + d^{k,k}(\varepsilon_k) -\textbf{x}^k(v_a)\textbf{1}^T \right)}{\varepsilon_k - v_a} = \det
\begin{pmatrix}
\sum_a \frac{1}{\varepsilon_k - v_a} & \textbf{1}^T \\
\sum_a \frac{\textbf{x}^k(v_a)}{\varepsilon_k - v_a} &  J^{k,k} + d^{k,k}(\varepsilon_k)
\end{pmatrix}.
\end{align}
The top left element is by definition $V_k$, and the remaining elements of the first column are seen to be
\begin{align}
\sum_a \frac{1}{(\varepsilon_i - v_a)(\varepsilon_k - v_a)} = -\frac{V_k - V_i}{\varepsilon_k - \varepsilon_i}.
\end{align}
Row operations reduce this determinant to a simple expression. Adding $\frac{1}{\varepsilon_k - \varepsilon_i}$ times the first row to each of $i$th other rows has three separate effects. First, each of the $i$th elements of the first column become $\frac{V_i}{\varepsilon_k - \varepsilon_i}$. Second, the diagonal elements of $J^{k,k} + d^{k,k}(\varepsilon_k)$  are ``repaired'': the factor $\frac{1}{\varepsilon_i - \varepsilon_k}$ is removed. Finally, the off-diagonal elements of $J^{k,k} + d^{k,k}(\varepsilon_k)$ are scaled by two factors. In particular, for $i,j\neq k$, the off-diagonal elements become
\begin{align}
\frac{1}{\varepsilon_i - \varepsilon_j} + \frac{1}{\varepsilon_k - \varepsilon_i} = \frac{(\varepsilon_k - \varepsilon_j)}{(\varepsilon_i - \varepsilon_j)(\varepsilon_k - \varepsilon_i)}.
\end{align}
A factor of $\frac{1}{\varepsilon_k - \varepsilon_i}$ can be removed from each of the $i$ rows (except the first), while a factor $(\varepsilon_k - \varepsilon_j)$ can be removed from each of the $j$ columns (except the first), and these factors cancel exactly. The first row becomes the $k$th row of $J$, the first column is the EBV $\{V\}$ (with the $k$th element in the first row), and the remaining $(N-1)\times (N-1)$ block is precisely $J^{k,k}$. The rows and columns may be reordered to yield the final expression
\begin{align}
\frac{1}{2} \braket{\{u\} | \hat{n}_k | \{v\}} = \eta \det J(k\rightarrow \textbf{V})
\end{align} 
where the matrix $J(k\rightarrow \textbf{V})$ is $J$ whose $k$th column has been replaced with the vector of EBV $\{V\}$. When the states are the same ($\{v\}=\{u\}$), and normalized,
\begin{align}
\gamma_k = \frac{\det \bar{J} (k\rightarrow \textbf{U})}{\det \bar{J}}
\end{align}
and Cramer's rule dictates that the 1-RDM elements $\gamma_k$ are the solutions of the linear equations
\begin{align} \label{eq:1rdm_cramer}
\bar{J} \bm{\gamma} = \textbf{U}.
\end{align}
If all one desires is the 1-RDM, then solving the linear equations \eqref{eq:1rdm_cramer} is the cleanest and most stable approach possible.

\subsection{Diagonal-correlation function}
The diagonal-correlation function is evaluated in essentially the same manner, though the intermediate summations are trickier, and ``repairing the damage'' to $J$ is \emph{much} more tedious. The final expression involves scaled second cofactors of the matrix $J$ rather than determinants differing by 2 columns. 

The diagonal-correlation function is evaluated as the sum of form factors \eqref{eq:d2d_ff}. The double form factors are obtained as the residues
\begin{align}
\braket{ \{u\} | S^+_k S^+_l | \{v\}_{a,b} } &= \lim_{v_a \rightarrow \varepsilon_k} \lim_{v_b \rightarrow \varepsilon_l} 
(v_a - \varepsilon_k) (v_b - \varepsilon_l) \braket{\{u\}|\{v\}} \\
&= \eta \det \left(J^{kl,kl} + d^{kl,kl}(\varepsilon_k) + d^{kl,kl}(\varepsilon_l) -d^{kl,kl}(v_a) - d^{kl,kl}(v_b) \right) 
\end{align}
where $J^{kl,kl}$ is $J$ without the $k$th and $l$th rows and the $k$th and $l$th columns etc. It is convenient to abbreviate the double sum as $I_{kl}$
\begin{align}
I_{kl} = \frac{1}{\eta} \sum_{a} \sum_{b (\neq a)} \frac{\braket{ \{u\} | S^+_k S^+_l | \{v\}_{a,b} }}{(\varepsilon_k - v_a)(\varepsilon_l - v_b)}.
\end{align}
The summation over $b$ will be performed first. Using lemma \ref{lem:mat_diag} gives
\begin{align}
I_{kl} = \sum_a \frac{1}{(\varepsilon_k - v_a)} \sum_{b (\neq a)} \frac{\det \left( J^{kl,kl} + d^{kl,kl}(\varepsilon_k) + d^{kl,kl}(\varepsilon_l) - d^{kl,kl}(v_a) - \textbf{x}^{kl}(v_b)\textbf{1}^T \right)}{(\varepsilon_l - v_b)} 
\end{align}
which, using lemma \ref{lem:update} gives
\begin{align}
I_{kl} = \sum_a \frac{1}{\varepsilon_k - v_a} \det
\begin{pmatrix}
\sum_{b (\neq a)} \frac{1}{\varepsilon_l - v_b} & \textbf{1}^T \\
\sum_{b (\neq a)} \frac{\textbf{x}^{kl}(v_b)}{\varepsilon_l - v_b} & J^{kl,kl} + d^{kl,kl}(\varepsilon_k) + d^{kl,kl}(\varepsilon_l) - d^{kl,kl}(v_a)
\end{pmatrix}.
\end{align}
The summations in the first column become
\begin{align}
\sum_{b (\neq a)} \frac{1}{\varepsilon_l - v_b} &= V_l - \frac{1}{\varepsilon_l - v_a} \\
\sum_{b (\neq a)} \frac{1}{\varepsilon_l - v_b} \frac{1}{\varepsilon_i - v_b} &= - \frac{V_l - V_i}{\varepsilon_l - \varepsilon_i} - \frac{1}{(\varepsilon_l - v_a)(\varepsilon_i - v_a)}.
\end{align}
The damage from $d^{kl,kl}(\varepsilon_l)$ may now be repaired as for the 1-RDM: add $\frac{1}{\varepsilon_l - \varepsilon_i}$ times the first row to each of the other $i$ rows, then factor $\frac{1}{\varepsilon_l - \varepsilon_i}$ from each row $i$ (except the first) and $(\varepsilon_l - \varepsilon_j)$ from each column $j$ (except the first). The result
\begin{align} \label{eq:Ikl_int}
I_{kl} = \sum_a \frac{1}{\varepsilon_k - v_a} \det \left(
\begin{array}{ccccc}
V_l - \frac{1}{\varepsilon_l - v_a} & \frac{1}{\varepsilon_l - \varepsilon_1} & \frac{1}{\varepsilon_l - \varepsilon_2} &  \dots & \frac{1}{\varepsilon_l - \varepsilon_N} \\
V_1 - \frac{1}{\varepsilon_1 - v_a} & \\
\vdots & \multicolumn{4}{c}{J^{kl,kl} + d^{kl,kl} (\varepsilon_k) - d^{kl,kl}(v_a) } \\
V_N - \frac{1}{\varepsilon_N - v_a}
\end{array} \right),
\end{align}
may be evaluated in more or less the same manner. The first row is the original $l$th row, and it is convenient to rearrange the rows to place them in the correct order. This permutation of the rows $\sigma_l$ introduces a sign. If $k<l$, the permutation $\sigma_l$ which places the $l$th row in the correct place may be accomplished with $l-2$ row swaps since the $k$th row is missing. If, on the other hand, $k$ is not less than $l$, then $l-1$ row swaps are required. Both situations are treated at once with $h(x)$ the Heaviside function
\begin{align}
h(x) = \begin{cases}
1 & x > 0 \\
0 & x \leq 0.
\end{cases}
\end{align}
The sign $| \sigma_l |$ is therefore $(-1)^{l-2 + h(k-l)}=(-1)^{l+h(k-l)}$. Keeping track of the intermediate signs is tedious, but manageable and necessary. The sum \eqref{eq:Ikl_int} becomes
\begin{align}
I_{kl} = (-1)^{l+h(k-l)} \sum_a \frac{1}{\varepsilon_k - v_a} \det \left( \begin{array}{c|c}
\textbf{V} - \frac{1}{ \bm{\varepsilon} - v_a} & J^{k,kl} + d^{k,kl}(\varepsilon_k) - d^{k,kl}(v_a)
\end{array} 
\right).
\end{align}
These matrices are of course $(N-1)\times (N-1)$ as the $k$th rows and columns are missing. To repeat the procedure, we will expand the determinants along the first column, so that
\begin{align}
I_{kl} = (-1)^{l+h(k-l)} \sum_a \sum_{i (\neq k)} \frac{(-1)^{i+1+h(i-k)}}{\varepsilon_k - v_a} \left( V_i - \frac{1}{\varepsilon_i - v_a} \right)
\det \left( J^{ki,kl} + d^{ki,kl}(\varepsilon_k) - d^{ki,kl}(v_a) \right).
\end{align}
Multiply each of the $i (\neq k)$ rows by $\frac{1}{\varepsilon_i - v_a}$ and each of the $j (\neq k,l)$ columns by $(\varepsilon_j - v_a)$ to arrive at
\begin{align}
I_{kl} &= \sum_a \frac{1}{\varepsilon_k - v_a} \left( V_l - \frac{1}{\varepsilon_l - v_a} \right)
\det \left( J^{kl,kl} + d^{kl,kl}(\varepsilon_k) - \textbf{x}^{kl}(v_a) \textbf{1}^T \right) \nonumber \\
&+ \sum_a \sum_{i (\neq k,l)} \frac{(-1)^{i+l+1+h(i-k)+h(k-l)}}{\varepsilon_k-v_a} \left(V_i - \frac{1}{\varepsilon_i - v_a} \right)
\frac{\varepsilon_l - v_a}{\varepsilon_i - v_a} \det \left( J^{ki,kl} + d^{ki,kl}(\varepsilon_k) -  \textbf{x}^{ki}(v_a) \textbf{1}^T  \right) \\
&= I^{kl}_{kl} + \sum_{i (\neq k,l)} (-1)^{i+l+1+h(i-k)+h(k-l)} I^{ki}_{kl} \label{eq:Ikl_int2}
\end{align}
since
\begin{align}
h(k-l) + h(l-k) = 1
\end{align}
and so
\begin{align}
(-1)^{l+h(k-l)+l+1+h(l-k)} = (-1)^{2l+2}=  1.
\end{align}
In \eqref{eq:Ikl_int2}, the $l$th term has been separated from the summation as it is special and so we will deal with it first. The summations over $a$ are performed separately, both using lemma \ref{lem:update} then repairing the damage with row operations. The first element of \eqref{eq:Ikl_int2} becomes
\begin{align}
I^{kl}_{kl} &= \sum_a \frac{1}{\varepsilon_k - v_a} \left( V_l - \frac{1}{\varepsilon_l - v_a} \right)
\det \left( J^{kl,kl} + d^{kl,kl}(\varepsilon_k) - \textbf{x}^{kl}(v_a) \textbf{1}^T \right) \\
&= \det
\begin{pmatrix}
K_{kl} & \textbf{1}^T \\
\textbf{w}^{kl} & J^{kl,kl} + d^{kl,kl}(\varepsilon_k)
\end{pmatrix},
\end{align}
where
\begin{align}
K_{kl} &= \sum_a \frac{1}{\varepsilon_k - v_a} \left( V_l - \frac{1}{\varepsilon_l - v_a} \right) \\
&= V_k V_l + \frac{V_k - V_l}{\varepsilon_k - \varepsilon_l} \\
&= \det \begin{pmatrix}
V_k + \frac{1}{\varepsilon_l - \varepsilon_k} & \frac{1}{\varepsilon_k - \varepsilon_l} \\
\frac{1}{\varepsilon_l - \varepsilon_k} & V_l + \frac{1}{\varepsilon_k - \varepsilon_l}
\end{pmatrix}
\end{align}
and the remaining elements of the first column are
\begin{align}
w^{kl}_i &= \sum_a \frac{1}{(\varepsilon_i - v_a)(\varepsilon_k - v_a)} \left( V_l - \frac{1}{(\varepsilon_l - v_a)} \right) \\
&= - V_l \frac{V_i - V_k}{(\varepsilon_i - \varepsilon_k)} - \frac{V_i}{(\varepsilon_i - \varepsilon_k)(\varepsilon_i - \varepsilon_l)}
- \frac{V_k}{(\varepsilon_k - \varepsilon_i)(\varepsilon_k - \varepsilon_l)} - \frac{V_l}{(\varepsilon_l - \varepsilon_i)(\varepsilon_l - \varepsilon_k)}.
\end{align}
To repair the damage, again add $\frac{1}{\varepsilon_k - \varepsilon_i}$ times the first row to each of the other $i$ rows, then factor $\frac{1}{\varepsilon_k - \varepsilon_i}$ from each row $i$ (except the first) and $(\varepsilon_k - \varepsilon_j)$ from each column $j$ (except the first), giving
\begin{align}
I^{kl}_{kl} &=  \det \left(
\begin{array}{ccccc}
K_{kl} & \frac{1}{\varepsilon_k - \varepsilon_1} &  \dots & \frac{1}{\varepsilon_k - \varepsilon_N} \\
K_{1l} & \\
\vdots & & J^{kl,kl} &  \\
K_{Nl} 
\end{array} \right).
\end{align}
The first row is now the $k$th row, which can be re-arranged with sign $(-1)^{k-2+h(l-k)}$, and the determinant can be expanded along the first column to yield
\begin{align}
I^{kl}_{kl} &= (-1)^{k-2+h(l-k)} \sum_{i(\neq l)} (-1)^{i+1 +h(i-l)} K_{il}\det J^{il,kl} \\
&= K_{kl} \det J^{kl,kl} + \sum_{i (\neq k,l)} (-1)^{k+i -1+ h(i-l)+h(l-k)} K_{il} \det J^{il,kl}.
\end{align}

The remaining summation elements $I^{ki}_{kl}$ of \eqref{eq:Ikl_int2} are evaluated in the same way, though it is substantially more tedious. First,
\begin{align}
I^{ki}_{kl} &= \sum_a \frac{(\varepsilon_l - v_a)}{(\varepsilon_k - v_a)(\varepsilon_i - v_a)} \det \left( J^{ki,kl} + d^{ki,kl}(\varepsilon_k) -  \textbf{x}^{ki}(v_a) \textbf{1}^T  \right) \\
&= \det
\begin{pmatrix}
\frac{\varepsilon_l - \varepsilon_k}{\varepsilon_i - \varepsilon_k} K_{ik} + \frac{\varepsilon_i - \varepsilon_l}{\varepsilon_i - \varepsilon_k} K_{ii} & \textbf{1}^T \\
\tilde{\textbf{w}}^{kl} & J^{ki,kl} + d^{ki,kl}(\varepsilon_k)
\end{pmatrix}
\end{align}
with 
\begin{align}
K_{ii} = V_i V_i + \frac{\partial V_i}{\partial \varepsilon_i}.
\end{align}
The $i$th row is missing, so the $l$th element of the vector $\tilde{w}^{kl}$ is
\begin{align}
\tilde{w}^{kl}_l = -\frac{1}{\varepsilon_i - \varepsilon_k}K_{ii} + \frac{1}{\varepsilon_i - \varepsilon_k} K_{ik}
\end{align}
while the other elements are 
\begin{align}
\tilde{w}^{kl}_{\alpha} = -\frac{(\varepsilon_i-\varepsilon_l)}{(\varepsilon_i - \varepsilon_k)(\varepsilon_i-\varepsilon_{\alpha})}K_{ii}
- \frac{(\varepsilon_k - \varepsilon_l)}{(\varepsilon_k - \varepsilon_i)(\varepsilon_k-\varepsilon_{\alpha})}K_{ik} 
- \frac{(\varepsilon_{\alpha}-\varepsilon_l)}{(\varepsilon_{\alpha}-\varepsilon_i)(\varepsilon_{\alpha}-\varepsilon_k)}K_{i\alpha}.
\end{align}
The damage is once again repaired by adding $\frac{1}{\varepsilon_k - \varepsilon_{\alpha}}$ times the first row to each of the other $\alpha$ rows, removing a factor of $\frac{1}{\varepsilon_k - \varepsilon_{\alpha}}$ from each row $\alpha$ (except the first) and $(\varepsilon_k - \varepsilon_{\beta})$ from each column $\beta$ (except the first). The resulting first column naturally splits into two, giving the result
\begin{align}
I^{ki}_{kl} &= \frac{\varepsilon_k - \varepsilon_i}{\varepsilon_k - \varepsilon_l} \det \begin{pmatrix}
\frac{\varepsilon_k-\varepsilon_l}{\varepsilon_k - \varepsilon_i} K_{ik} & \frac{1}{\varepsilon_k - \varepsilon_{\bm{\beta}}} \\
\frac{\varepsilon_{\bm{\alpha}}-\varepsilon_l}{\varepsilon_{\bm{\alpha}}-\varepsilon_i}K_{i\bm{\alpha}} & J^{ki,kl}
\end{pmatrix}
+\frac{(\varepsilon_i-\varepsilon_l)(\varepsilon_i - \varepsilon_k)}{(\varepsilon_k-\varepsilon_l)} K_{ii} \det (J^i_i | J^{i,l}).
\end{align}
Note that in the $l$th row of the first determinant, the element in the first column is zero. Here $J^i_i$ is the $i$th column of $J$ without the $i$th row, while $J^{i,l}$ is $J$ without the $i$th row and $l$th column. This last determinant has a repeated $i$th column and is thus zero.

The rows may be reordered (again giving a factor of $(-1)^{k-2+h(i-k)}$) and expanded along the first column:
\begin{align}
I^{ki}_{kl} &= \frac{\varepsilon_k - \varepsilon_i}{\varepsilon_k - \varepsilon_l} (-1)^{k-2+h(i-k)} \sum_{j (\neq i, l)} (-1)^{j+1+h(j-i)} 
\frac{\varepsilon_j - \varepsilon_l}{\varepsilon_j - \varepsilon_i} K_{ij} \det J^{ij,kl} \\
&= K_{ik} \det J^{ik,kl} + \sum_{j (\neq i, k,l)} (-1)^{k+j-1+h(i-k)+h(j-i)}\frac{(\varepsilon_k - \varepsilon_i)(\varepsilon_j - \varepsilon_l)}{(\varepsilon_k-\varepsilon_l)(\varepsilon_j - \varepsilon_i)} K_{ij} \det J^{ij,kl}
\end{align}
The summation is restricted over $i$ and $l$ as the $i$th row is missing, and the coefficient of the $l$th row is zero.

Now, all the tedious sign-tracking pays off. With the second cofactors of the matrix $J$
\begin{align}
A[J]^{ij,kl} = (-1)^{i+j+k+l+h(i-j)+h(k-l)} \det J^{ij,kl}
\end{align}
the final summations are simple. In particular,
\begin{align}
\frac{1}{\eta} \frac{1}{4} \braket{\{u\}|\hat{n}_k\hat{n}_l|\{v\}} &= K_{kl} A[J]^{kl,kl} + \sum_{i (\neq k,l)} K_{il} A[J]^{il,kl} + \sum_{i (\neq k,l)} K_{ik} A[J]^{ki,kl} \nonumber \\
&+ \sum_{i < j (\neq k,l)} \frac{(\varepsilon_k - \varepsilon_i)(\varepsilon_l-\varepsilon_j)+(\varepsilon_k - \varepsilon_j)(\varepsilon_l-\varepsilon_i)}{(\varepsilon_k-\varepsilon_l)(\varepsilon_j-\varepsilon_i)} K_{ij} A[J]^{ij,kl}.
\end{align}
Like the case for rapidities, a clean result is obtained: a sum over second cofactors of a common matrix. The simplification to the 1-RDM case is obvious: expansion in first cofactors. The generalization to higher $k$-RDMs is simple: expansion in terms of rank-$k$ cofactors. 

\subsection{Pair-correlation function}

It is convenient to define
\begin{align}
I^{(1)}_{kl} &= \frac{1}{\eta} \sum_a \frac{\braket{\{u\} | S^+_k | \{v\}_a}}{v_a - \varepsilon_l} \\
I^{(2)}_{kl} &= \frac{1}{\eta} \sum_a \sum_{b (\neq a)} \frac{\braket{\{u\}|S^+_k S^+_l |\{v\}}}{(v_a-\varepsilon_l)(v_b - \varepsilon_l)}
\end{align}
so that 
\begin{align}
\frac{1}{\eta} \braket{\{u\}|S^+_k S^-_l | \{v\}} = I^{(1)}_{kl} - I^{(2)}_{kl}.
\end{align}
The two summations will be performed independently. The single summation is evaluated in the same way as for the 1-RDM until
\begin{align}
I^{(1)}_{kl}
= \sum_a \frac{\det \left(J^{k,k} + d^k(\varepsilon_k) -\textbf{x}^k(v_a)\textbf{1}^T \right)}{\varepsilon_l - v_a}
\end{align}
which again simplifies to a single determinant of rank $N$, though the first column is different: the first element is $V_l$ (rather than $V_k$), the $l$th element becomes
\begin{align}
\sum_a \frac{1}{(\varepsilon_l - v_a)^2} := -\frac{\partial V_l}{\partial \varepsilon_l},
\end{align}
while the other elements remain (for $i\neq k,l$), $-\frac{V_l-V_i}{\varepsilon_l - \varepsilon_i}$. The damage to the rest of $J$ will be repaired in the same manner as before, by first adding $\frac{1}{\varepsilon_k - \varepsilon_i}$ times the first row to each of the $i$ rows, then by again factoring $\frac{1}{\varepsilon_k - \varepsilon_i}$ from each row (except the first) and $(\varepsilon_k - \varepsilon_j)$ from each column (except the first). The rows and columns may be interchanged to yield the determinant of $J$ with the $k$th column replaced by the sum of two vectors $\textbf{p}^k + \textbf{q}^k$, or using linearity in the columns,
\begin{align}
I^{(1)}_{kl} = \det J(k \rightarrow \textbf{p}^k) + \det J(k \rightarrow \textbf{q}^k)
\end{align}
where
\begin{align}
p^k_i = \begin{cases}
0, & i = k \\
V_l - (\varepsilon_k - \varepsilon_l) \frac{\partial V_l}{\partial \varepsilon_l}, & i = l \\
\frac{(\varepsilon_i - \varepsilon_k)}{(\varepsilon_i - \varepsilon_l)} V_i, & i\neq k,l
\end{cases}
\end{align}
and
\begin{align}
q^k_i = \begin{cases}
0, & i = l \\
\frac{(\varepsilon_k - \varepsilon_l)}{(\varepsilon_i - \varepsilon_l)}V_l, & i \neq l
\end{cases}.
\end{align}
Rearranging the rows and columns here never introduces a sign as both rows and columns must be interchanged the same number of times. Notice that except for the $l$th element, $\textbf{q}^k$ is $(\varepsilon_k - \varepsilon_l)V_l$ times the $l$th column. Therefore the $l$th column can be subtracted from the $k$th, and the the only non-zero contribution is
\begin{align}
\det J (k \rightarrow \textbf{q}^k) = (-1)^{k+l} (\varepsilon_l - \varepsilon_k) V_l J_{ll} \det J^{l,k}
\end{align} 
where $J^{l,k}$ is the matrix obtained from J by removing the $l$th row and $k$th column. The other determinant is expanded along the $k$th column to give
\begin{align}
\det J(k\rightarrow \textbf{p}^k) = (-1)^{k+l}\left( V_l - (\varepsilon_k - \varepsilon_l) \frac{\partial V_l}{\partial \varepsilon_l} \right) \det J^{l,k} + \sum_{i (\neq k,l)} (-1)^{i+k} \frac{(\varepsilon_i - \varepsilon_k)}{(\varepsilon_i - \varepsilon_l)} V_i \det J^{i,k}.
\end{align}
The single summation is thus
\begin{align}
I^{(1)}_{kl} =
\left( V_l - (\varepsilon_k - \varepsilon_l) \left(\frac{\partial V_l}{\partial \varepsilon_l} + V_l J_{ll} \right) \right) A[J]^{l,k} 
+ \sum_{i (\neq k,l)} \frac{(\varepsilon_i - \varepsilon_k)}{(\varepsilon_i - \varepsilon_l)} V_i A[J]^{i,k}.
\end{align}

The double summation is evaluated in a similar manner as for the diagonal-correlation function, so we will skip a few of the intermediate steps. We can again perform the summation over $b$ first, and proceed directly to
\begin{align}
I^{(2)}_{kl} = \sum_a \frac{1}{\varepsilon_l - v_a} (-1)^{l + h(k-l)} \det \left( \begin{array}{c|c}
\textbf{V} - \frac{1}{ \bm{\varepsilon} - v_a} & J^{k,kl} + d^{k,kl}(\varepsilon_k) - d^{k,kl}(v_a)
\end{array}   \right).
\end{align}
Next, scale the rows and columns to make it a rank-1 update, expand along the first column, collect the results with lemma \ref{lem:update} and repair the damage by adding $\frac{1}{\varepsilon_k - \varepsilon_{\alpha}}$ times the first row to each of the $\alpha$ rows (and scaling the rows and columns) to get
\begin{align} \label{eq:i2_int}
I^{(2)}_{kl} = (-1)^{l + h(k-l)} \sum_{i (\neq k)} (-1)^{i+1+h(i-k)} \frac{\varepsilon_k - \varepsilon_i}{\varepsilon_k - \varepsilon_l}
\det \left(
\begin{array}{ccccc}
K_{ii} & \frac{1}{\varepsilon_k - \varepsilon_1} &  \dots & \frac{1}{\varepsilon_k - \varepsilon_N} \\
\frac{\varepsilon_k - \varepsilon_1}{\varepsilon_i - \varepsilon_1} K_{i1} + \frac{\varepsilon_k - \varepsilon_i}{\varepsilon_1 - \varepsilon_i}K_{ii} & \\
\vdots & & J^{ik,kl} &  \\
\frac{\varepsilon_k - \varepsilon_N}{\varepsilon_i - \varepsilon_N} K_{iN} + \frac{\varepsilon_k - \varepsilon_i}{\varepsilon_N - \varepsilon_i}K_{ii}
\end{array} \right).
\end{align}
These determinants can be split into two, by grouping the contributions from $K_{ii}$. For $i \neq l$, determinants proportional to $K_{ii}$ will vanish as their first column is a scalar multiple of the $i$th column. For $i=l$, the $l$th column is missing, and the final contribution is $(\varepsilon_l - \varepsilon_k) K_{ll} A[J]^{l,k}$. The only difficulty is keeping track of the signs: the $k$th row is in the first row, so must be reordered with sign $(-1)^{k-2+h(l-k)}$, the $l$th column is in the first column, so is reordered with sign $(-1)^{l-2+h(k-l)}$. Combining these with the signs in \eqref{eq:i2_int} gives
\begin{align}
(-1)^{l+h(k-l)}(-1)^{l+1+h(l-k)}(-1)^{k+h(l-k)}(-1)^{l+h(k-l)} = (-1)^{k+l+1}.
\end{align}
The extra sign is given to $-(\varepsilon_k - \varepsilon_l) = (\varepsilon_l - \varepsilon_k)$, so that only $(-1)^{k+l}$ remains for $A[J]^{l,k} = (-1)^{k+l}\det J^{l,k}$. The rest of $I^{(2)}_{kl}$, which contains the determinants with columns not proportional to $K_{ii}$, 
\begin{align}
I^{(2)'}_{kl} = (-1)^{l + h(k-l)} \sum_{i (\neq k)} (-1)^{i+1+h(i-k)} \frac{\varepsilon_k - \varepsilon_i}{\varepsilon_k - \varepsilon_l}
\det \left(
\begin{array}{ccccc}
0 & \frac{1}{\varepsilon_k - \varepsilon_1} &  \dots & \frac{1}{\varepsilon_k - \varepsilon_N} \\
\frac{\varepsilon_k - \varepsilon_1}{\varepsilon_i - \varepsilon_1} K_{i1}  & \\
\vdots & & J^{ik,kl} &  \\
\frac{\varepsilon_k - \varepsilon_N}{\varepsilon_i - \varepsilon_N} K_{iN} 
\end{array} \right)
\end{align}
is easily evaluated. The rows are rearranged, giving a sign $(-1)^{k+h(i-k)}$, and the determinants are expanded along the first column
\begin{align}
I^{(2)'}_{kl} = (-1)^{l + h(k-l)} \sum_{i (\neq k)} (-1)^{i+1+k+2h(i-k)} \frac{\varepsilon_k - \varepsilon_i}{\varepsilon_k - \varepsilon_l} \sum_{\alpha (\neq i,k)}
(-1)^{\alpha + 1 + h(\alpha - i)} \frac{\varepsilon_k - \varepsilon_{\alpha}}{\varepsilon_i - \varepsilon_{\alpha}}K_{i\alpha} \det J^{\alpha i,kl}.
\end{align}
Collecting the signs gives $(-1)^{i + \alpha + k + l + h(\alpha -i)+h(k-l)}$, the sign desired in the definition of the second cofactor, so that finally $I^{(2)}_{kl}$ becomes
\begin{align}
I^{(2)}_{kl} = (\varepsilon_l - \varepsilon_k)K_{ll} A[J]^{l,k} + 2 \sum_{i (\neq k,l)} \frac{\varepsilon_k - \varepsilon_i}{\varepsilon_l - \varepsilon_i} K_{i l}
A[J]^{i l,kl} 
+ 2\sum_{\substack{ i < j  \\ (\neq k,l) } } \frac{(\varepsilon_k - \varepsilon_i)(\varepsilon_k - \varepsilon_j)}{(\varepsilon_k - \varepsilon_l)(\varepsilon_j-\varepsilon_i)}
K_{ij} A[J]^{ij,kl}.
\end{align}

In combining $I^{(1)}_{kl}$ and $I^{(2)}_{kl}$, only the coefficient of $A[J]^{l,k}$ is modified
\begin{align}
\frac{1}{\eta} \braket{\{u\}|S^+_k S^-_l | \{v\}} &= \left( V_l + (\varepsilon_k - \varepsilon_l)(V_l V_l - V_l J_{ll}) \right) A[J]^{l,k} + \sum_{i (\neq k,l)}\frac{\varepsilon_i-\varepsilon_k}{\varepsilon_i-\varepsilon_l} V_i A[J]^{i,k} \nonumber \\
&- 2 \sum_{i (\neq k,l)} \frac{\varepsilon_k - \varepsilon_i}{\varepsilon_l - \varepsilon_i} K_{i l}
A[J]^{i l,kl} 
- 2\sum_{\substack{ i < j  \\ (\neq k,l) } } \frac{(\varepsilon_k - \varepsilon_i)(\varepsilon_k - \varepsilon_j)}{(\varepsilon_k - \varepsilon_l)(\varepsilon_j-\varepsilon_i)}
K_{ij} A[J]^{ij,kl}.
\end{align}

\subsection{Reduced Density Matrix Final expressions}
We have seen that to compute density matrix elements in terms of the EBV, all that is required are the first and second cofactors of the matrix $J$. Evaluated directly, this would be quite expensive as there are $\mathcal{O}(N^4)$ second cofactors and each would require $\mathcal{O}(N^3)$ floating point operations. As was the case for the rapidity-based expressions, elementary results of linear algebra reduce this cost dramatically. 

First, \emph{scaled} second cofactors are computable with scaled first cofactors by Jacobi's theorem\cite{vein_book}
\begin{align}
\frac{A[\bar{J}]^{ij,kl}}{\det \bar{J}} = \frac{A[\bar{J}]^{i,k}}{\det \bar{J}} \frac{A[\bar{J}]^{j,l}}{\det \bar{J}}-\frac{A[\bar{J}]^{i,l}}{\det \bar{J}}\frac{A[\bar{J}]^{j,k}}{\det \bar{J}},
\end{align}
which appear naturally when normalizing the RDM elements. Jacobi's theorem holds to any order: $k$th-order scaled cofactors are $k\times k$ determinants of first scaled cofactors.

Second, the matrix inverse of $\bar{J}$ may be written as its adjugate divided by its determinant
\begin{align}
\bar{J}^{-1} = \frac{\text{adj}(\bar{J})}{\det \bar{J}},
\end{align}
where the adjugate matrix $\text{adj}(\bar{J})$ is the transpose of the matrix of cofactors
\begin{align}
\text{adj}(\bar{J})_{ij} = A[\bar{J}]^{j,i}.
\end{align}
The scaled first cofactors are therefore obtained directly as the elements of the transpose of $\bar{J}^{-1}$. The inverse is computed numerically with $\mathcal{O}(N^3)$ cost, and the 2-RDM may constructed with $\mathcal{O}(N^4)$ cost: there are $\mathcal{O}(N^2)$ elements, and each requires computing a double sum. Notice that this is nearly the same scaling as for the rapidity expressions $\mathcal{O}(N^2 M^2)$. Unless $N >> M$, any benefit of using rapidities to compute the 2-RDM would be mitigated by the cost of computing the rapidities and the potential for loss of numerical precision.

The final expressions are
\begin{align} \label{eq:ebv_1dm}
\gamma_{k} = \sum_l U_l \bar{J}^{-1}_{kl}
\end{align}

\begin{align} \label{eq:ebv_d2d}
D_{kl} &= K_{kl} (\bar{J}^{-1}_{kk}\bar{J}^{-1}_{ll}-\bar{J}^{-1}_{lk}\bar{J}^{-1}_{kl}) 
+ \sum_{j (\neq l)} K_{jk} (\bar{J}^{-1}_{kk}\bar{J}^{-1}_{lj}-\bar{J}^{-1}_{lk}\bar{J}^{-1}_{kj}) 
+ \sum_{j (\neq k)} K_{jl} (\bar{J}^{-1}_{kj}\bar{J}^{-1}_{ll}-\bar{J}^{-1}_{lj}\bar{J}^{-1}_{kl})  \nonumber \\
&+ \sum_{\substack{ i < j  \\ (\neq k,l) } } \frac{(\varepsilon_k - \varepsilon_i)(\varepsilon_l - \varepsilon_j)+(\varepsilon_k - \varepsilon_j)(\varepsilon_l - \varepsilon_i)}{(\varepsilon_k - \varepsilon_l)(\varepsilon_j - \varepsilon_i)} K_{ij} (\bar{J}^{-1}_{ki}\bar{J}^{-1}_{lj}-\bar{J}^{-1}_{li}\bar{J}^{-1}_{kj}) 
\end{align}

\begin{align} \label{eq:ebv_d2p}
P_{kl} &= \left(2 U_l + \sum_{i (\neq k,l)} \frac{\varepsilon_i - \varepsilon_k}{\varepsilon_i - \varepsilon_l} U_i - \frac{2M}{g} \right) \bar{J}^{-1}_{kl}  \nonumber \\
&+ \sum_{i (\neq k,l)} \frac{\varepsilon_i - \varepsilon_k}{\varepsilon_i - \varepsilon_l} (U_i \bar{J}^{-1}_{ki} -2 K_{il} (\bar{J}^{-1}_{ki}\bar{J}^{-1}_{ll}-\bar{J}^{-1}_{li}\bar{J}^{-1}_{kl}) ) \nonumber \\
&-2 \sum_{\substack{ i < j  \\ (\neq k,l) } } \frac{(\varepsilon_k - \varepsilon_i)(\varepsilon_k - \varepsilon_j)}{(\varepsilon_k - \varepsilon_l)(\varepsilon_j - \varepsilon_i)}
K_{ij} (\bar{J}^{-1}_{ki}\bar{J}^{-1}_{lj}-\bar{J}^{-1}_{li}\bar{J}^{-1}_{kj}).
\end{align}
Density matrix elements between distinct on-shell RG states, transition density matrix elements (TDM), are not directly computable from a matrix inverse, as the corresponding matrix $J$ would be singular. Cofactors of a singular matrix are not independent, so it would not be required to compute them all. We are currently exploring better strategies to compute the TDM elements and will report our results in a future contribution.

\subsection{Derivatives of RDM elements}
Minimization of the energy functional \eqref{eq:e_functional} should benefit from exact first-, and if possible, second-derivatives with respect to the variational parameters $\{\varepsilon\}$ and $g$. These are evaluated directly from derivatives of $\gamma_k$, $D_{kl}$ and $P_{kl}$. All of the required intermediates are obtained as linear equations with the matrix $\bar{J}$ or derivatives of the matrix $\bar{J}$. 

Differentiating \eqref{eq:ebv_1dm} is simple, 
\begin{align} 
\frac{\partial \gamma_k}{\partial \varepsilon_m} = \sum_l \left( \frac{\partial U_l}{\partial \varepsilon_m} \bar{J}^{-1}_{kl} + U_l \frac{\partial \bar{J}^{-1}_{kl}}{\partial \varepsilon_m} \right)
\end{align}
while \eqref{eq:ebv_d2d} and \eqref{eq:ebv_d2p} are tedious but straightforward. To be clear, the symbol $\frac{\partial \bar{J}^{-1}_{kl}}{\partial \varepsilon_m}$ should be understood as the $(k,l)-$th element of $\frac{\partial \bar{J}^{-1}}{\partial \varepsilon_m}$. The derivatives of the diagonal-correlation function are
\begin{align} \label{eq:der_d2d}
\frac{\partial D_{kl}}{\partial \varepsilon_m} &= 
\frac{\partial K_{kl}}{\partial \varepsilon_m}   \frac{A[\bar{J}]^{kl,kl}}{\det \bar{J}} 
+ K_{kl} \frac{\partial}{\partial \varepsilon_m} \left( \frac{A[\bar{J}]^{kl,kl}}{\det \bar{J}} \right) \nonumber \\
&+ \sum_{j (\neq l)}  \frac{\partial K_{jk}}{\partial \varepsilon_m} \frac{A[\bar{J}]^{il,kl}}{\det \bar{J}} 
+ K_{jk} \frac{\partial}{\partial \varepsilon_m} \left( \frac{A[\bar{J}]^{il,kl}}{\det \bar{J}} \right) \nonumber \\
&+ \sum_{j (\neq k)} \frac{\partial K_{jl}}{\partial \varepsilon_m} \frac{A[\bar{J}]^{ki,kl}}{\det \bar{J}} 
+ K_{jl} \frac{\partial}{\partial \varepsilon_m} \left( \frac{A[\bar{J}]^{ki,kl}}{\det \bar{J}} \right) \nonumber \\
&+ \sum_{\substack{ i < j  \\ (\neq k,l) } } \frac{\partial T^D_{ijkl}}{\partial \varepsilon_m}   K_{ij} \frac{A[\bar{J}]^{ij,kl}}{\det \bar{J}} 
+ T^D_{ijkl}  \frac{\partial K_{ij}}{\partial \varepsilon_m} \frac{A[\bar{J}]^{ij,kl}}{\det \bar{J}} 
+ T^D_{ijkl}  K_{ij} \frac{\partial}{\partial \varepsilon_m} \left( \frac{A[\bar{J}]^{ij,kl}}{\det \bar{J}} \right)
\end{align}
with the derivatives of $K_{ij}$,
\begin{align}
\frac{\partial K_{ij}}{\partial \varepsilon_m} = \frac{\partial U_i}{\partial \varepsilon_m}U_j + U_i \frac{\partial U_j}{\partial \varepsilon_m} 
+ \frac{1}{\varepsilon_i - \varepsilon_j} \left( \frac{\partial U_i}{\partial \varepsilon_m} - \frac{\partial U_j}{\partial \varepsilon_m} \right)
- \frac{U_i - U_j}{(\varepsilon_i - \varepsilon_j)^2}(\delta_{im}-\delta_{jm})
\end{align}
and the derivatives of the second cofactors
\begin{align}
\frac{\partial}{\partial \varepsilon_m} \left( \frac{A[\bar{J}]^{ij,kl}}{\det \bar{J}} \right) = 
\frac{\partial \bar{J}^{-1}_{ki}}{\partial \varepsilon_m} \bar{J}^{-1}_{lj}
+ \bar{J}^{-1}_{ki} \frac{\partial \bar{J}^{-1}_{lj}}{\partial \varepsilon_m}
- \frac{\partial \bar{J}^{-1}_{li}}{\partial \varepsilon_m} \bar{J}^{-1}_{kj}
- \bar{J}^{-1}_{li} \frac{\partial \bar{J}^{-1}_{kj}}{\partial \varepsilon_m}.
\end{align}
In \eqref{eq:der_d2d}, we have used
\begin{align}
T^D_{ijkl} &= \frac{(\varepsilon_k - \varepsilon_i)(\varepsilon_l - \varepsilon_j)+(\varepsilon_k - \varepsilon_j)(\varepsilon_l - \varepsilon_i)}{(\varepsilon_k - \varepsilon_l)(\varepsilon_j - \varepsilon_i)} \\
\frac{\partial T^D_{ijkl}}{\partial \varepsilon_m}&=
\frac{(\delta_{km}-\delta_{im})(\varepsilon_l-\varepsilon_j)+(\varepsilon_k - \varepsilon_i)(\delta_{lm}-\delta_{jm})+(\delta_{km}-\delta_{jm})(\varepsilon_l-\varepsilon_i)+(\varepsilon_k - \varepsilon_j)(\delta_{lm}-\delta_{im})}{(\varepsilon_k - \varepsilon_l)(\varepsilon_j - \varepsilon_i)} \nonumber \\
&- \frac{((\varepsilon_k-\varepsilon_i)(\varepsilon_l-\varepsilon_j)+(\varepsilon_k-\varepsilon_j)(\varepsilon_l-\varepsilon_i)) ((\delta_{km}-\delta_{lm})(\varepsilon_j-\varepsilon_i)+(\varepsilon_k-\varepsilon_l)(\delta_{jm}-\delta_{im}))}{(\varepsilon_k - \varepsilon_l)^2(\varepsilon_j-\varepsilon_i)^2}.
\end{align}
Likewise,
\begin{align}
\frac{\partial P_{kl}}{\partial \varepsilon_m} &=
\left(2 \frac{\partial U_l}{\partial \varepsilon_m} + \sum_{i (\neq k,l)} \left( \frac{\partial t^P_{ikl}}{\partial \varepsilon_m}  U_i + t^P_{ikl} \frac{\partial U_i}{\partial \varepsilon_m} \right) \right) \bar{J}^{-1}_{kl} 
+ \left(2 U_l + \sum_{i (\neq k,l)} t^P_{ikl}  U_i - \frac{2M}{g} \right) \frac{\partial \bar{J}^{-1}_{kl}}{\partial \varepsilon_m}  \nonumber \\
&+ \sum_{i (\neq k,l)} \frac{\partial t^P_{ikl}}{\partial \varepsilon_m} \left( U_i \bar{J}^{-1}_{ki} -2 K_{il} \frac{A[\bar{J}]^{il,kl}}{\det \bar{J}} \right) \nonumber \\
&+ \sum_{i (\neq k,l)} t^P_{ikl} \left( \frac{\partial U_i}{\partial\varepsilon_m} \bar{J}^{-1}_{ki} + U_i \frac{\partial \bar{J}^{-1}_{ki}}{\partial \varepsilon_m} -2 \frac{\partial K_{il}}{\partial \varepsilon_m} \frac{A[\bar{J}]^{il,kl}}{\det \bar{J}} - 2 K_{il} \frac{\partial}{\partial \varepsilon_m}\left( \frac{A[\bar{J}]^{il,kl}}{\det \bar{J}} \right)  \right) \nonumber \\
&-2 \sum_{\substack{ i < j  \\ (\neq k,l) } } \left( \frac{\partial T^P_{ijkl}}{\partial \varepsilon_m} K_{ij} \frac{A[\bar{J}]^{ij,kl}}{\det \bar{J}} 
+  T^P_{ijkl} \frac{\partial K_{ij}}{\partial \varepsilon_m} \frac{A[\bar{J}]^{ij,kl}}{\det \bar{J}}
+ T^P_{ijkl} K_{ij} \frac{\partial}{\partial \varepsilon_m}\left( \frac{A[\bar{J}]^{ij,kl}}{\det \bar{J}}\right) \right)
\end{align}
with
\begin{align}
t^P_{ikl}  &= \frac{\varepsilon_i - \varepsilon_k}{\varepsilon_i - \varepsilon_l} \\
\frac{\partial t^P_{ikl}}{\partial \varepsilon_m} &= \frac{(\delta_{im}-\delta_{km})(\varepsilon_i-\varepsilon_l)-(\varepsilon_i-\varepsilon_k)(\delta_{im}-\delta_{lm})}{(\varepsilon_i-\varepsilon_l)^2} \\
T^P_{ijkl} &= \frac{(\varepsilon_k - \varepsilon_i)(\varepsilon_k - \varepsilon_j)}{(\varepsilon_k - \varepsilon_l)(\varepsilon_j - \varepsilon_i)} \\
\frac{\partial T^P_{ijkl}}{\partial \varepsilon_m} &= 
\frac{(\delta_{km}-\delta_{im})(\varepsilon_k-\varepsilon_j)+(\varepsilon_k-\varepsilon_i)(\delta_{km}-\delta_{jm})}{(\varepsilon_k-\varepsilon_l)(\varepsilon_j-\varepsilon_i)} \nonumber \\
&- \frac{(\varepsilon_k-\varepsilon_i)(\varepsilon_k-\varepsilon_j) ( (\delta_{km}-\delta_{lm})(\varepsilon_j-\varepsilon_i)+(\varepsilon_k-\varepsilon_l)(\delta_{jm}-\delta_{im}) )}{(\varepsilon_k-\varepsilon_l)^2(\varepsilon_j-\varepsilon_i)^2}
.
\end{align}
Similar expressions are obtained for the derivatives with respect to $g$, though they are simpler as
\begin{align}
\frac{\partial T^D_{ijkl}}{\partial g}=\frac{\partial t^P_{ikl}}{\partial g}=\frac{\partial T^P_{ijkl}}{\partial g}=0.
\end{align}
Symbolic derivatives of $\bar{J}^{-1}$ with respect to $\{\varepsilon\}$ and $g$ would be quite complicated in general, but again, elementary results of linear algebra save the day. For $C(z)$ a matrix dependant upon $z$, the derivative of the inverse $C^{-1}$ is
\begin{align} \label{eq:inv-dev}
\frac{\partial C^{-1}}{\partial z} = - C^{-1} \frac{\partial C(z)}{\partial z} C^{-1}.
\end{align}
Usually the derivative $\frac{\partial C(z)}{\partial z}$ is straightforward to compute. The relevant derivatives of $\bar{J}$ are easily evaluated and \emph{sparse}: $\frac{\partial \bar{J}}{\partial \varepsilon_m}$ is diagonal along with one non-zero row and one non-zero column
\begin{align}
\left(\frac{\partial \bar{J}}{\partial \varepsilon_m}\right)_{ij} = 
\begin{cases}
2\frac{\partial U_i}{\partial \varepsilon_m} + \sum_{k(\neq i)}\frac{1}{(\varepsilon_k - \varepsilon_m)^2}, & i = j = m \\
2\frac{\partial U_i}{\partial \varepsilon_m} - \frac{1}{(\varepsilon_m - \varepsilon_i)^2}, & i=j\neq m \\
-\frac{1}{(\varepsilon_m - \varepsilon_j)^2}, & i=m\neq j \\
\frac{1}{(\varepsilon_i - \varepsilon_m)^2}, & i\neq j=m
\end{cases}
\end{align}
while $\frac{\partial \bar{J}}{\partial g}$ is diagonal
\begin{align}
\left( \frac{\partial \bar{J}}{\partial g} \right)_{ii} = 2\frac{\partial U_i}{\partial g} + \frac{2}{g^2}.
\end{align}
Derivatives of $\bar{J}^{-1}$ are then obtained from \eqref{eq:inv-dev}, requiring only two matrix multiplications each. 

The derivatives of the EBV with respect to single particle energies
\begin{align}
\frac{\partial \textbf{U}}{\partial \bm{\varepsilon}} = \begin{pmatrix}
\frac{\partial U_1}{\partial \varepsilon_1} & \dots & \frac{\partial U_1}{\partial \varepsilon_N} \\
\vdots & \ddots & \vdots \\
\frac{\partial U_N}{\partial \varepsilon_1} & \dots & \frac{\partial U_N}{\partial \varepsilon_N}
\end{pmatrix}
\end{align}
are solutions of linear equations
\begin{align} \label{eq:ebv_der_lineq}
\bar{J} \frac{\partial \textbf{U}}{\partial \bm{\varepsilon}} = B
\end{align}
where the RHS is
\begin{align}
B_{ij} = \begin{cases}
\sum_{k (\neq i)} \frac{U_k - U_i}{(\varepsilon_k - \varepsilon_i)^2}, & i=j \\
\frac{U_i-U_j}{(\varepsilon_i - \varepsilon_j)^2}, & i \neq j.
\end{cases}
\end{align}
These equations are solved in $\mathcal{O}(N^3)$ operations, either by pLU decomposing $\bar{J}$ and solving the $N$ sets of equations, or since we must already compute $\bar{J}^{-1}$, just performing the multiplication $\frac{\partial \textbf{U}}{\partial \bm{\varepsilon}} = \bar{J}^{-1} B$. The same approach gives derivatives with respect to $g$:
\begin{align}
\bar{J} \frac{\partial \textbf{U}}{\partial g} = \textbf{b}^g
\end{align}
where
\begin{align}
b^g_i = -\frac{2U_i}{g^2}.
\end{align}

With these ingredients, first derivatives of $\gamma_k$, $D_{kl}$ and $P_{kl}$ are computed, and hence the gradient of the energy functional \eqref{eq:e_functional} is easily evaluated. The Hessian of \eqref{eq:e_functional} may be evaluated in exactly the same manner, though we will not present the details. Second, and higher, derivatives of $\bar{J}^{-1}$ are evaluated by iterating the first derivative expressions, e.g.
\begin{align}
\frac{\partial^2 \bar{J}^{-1}}{\partial \varepsilon_m \partial \varepsilon_n } = 
\bar{J}^{-1} \frac{\partial \bar{J}}{\partial \varepsilon_n} \bar{J}^{-1} \frac{\partial \bar{J}}{\partial \varepsilon_m} \bar{J}^{-1}
+ \bar{J}^{-1} \frac{\partial \bar{J}}{\partial \varepsilon_m} \bar{J}^{-1} \frac{\partial \bar{J}}{\partial \varepsilon_n} \bar{J}^{-1}
- \bar{J}^{-1} \frac{\partial^2 \bar{J}}{\partial \varepsilon_m \partial \varepsilon_n} \bar{J}^{-1}.
\end{align}
Second derivatives of the EBV with respect to $\{\varepsilon\}$ and $g$ are again obtained as solutions of linear equations \emph{with the same matrix} $\bar{J}$ as \eqref{eq:ebv_der_lineq}, but with different right hand sides, e.g.
\begin{align}
\bar{J} \frac{\partial}{\partial \varepsilon_m} \frac{\partial \textbf{U}}{\partial \bm{\varepsilon}} =
- \frac{\partial \bar{J}}{\partial \varepsilon_m} \frac{\partial \textbf{U}}{\partial \bm{\varepsilon}} + \frac{\partial B}{\partial \varepsilon_m}
\end{align}
which is most direct as both $\frac{\partial \textbf{U}}{\partial \bm{\varepsilon}}$ and $\frac{\partial \bar{J}}{\partial \varepsilon_m}$ must be computed for the gradient. All the intermediate elements required to compute the Hessian are thus obtained as derivatives of $\bar{J}$, along with linear systems with $\bar{J}$.

\section{Numerical Stability} \label{sec:stability}
As numerically inverting matrices is usually avoided, in this section we demonstrate that it is acceptable for the matrix $\bar{J}$ so long as the single-particle energies $\{\varepsilon\}$ are non-degenerate. We solve the EBV equations with the method outlined in refs.\cite{faribault:2011,elaraby:2012,fecteau:2022} It is convenient to rescale the EBV by the interaction $g$
\begin{align}
\tilde{U}_i = gU_i
\end{align}
which modifies the EBV equations to
\begin{align}
\tilde{U}^2_i - 2\tilde{U}_i - g \sum_{k (\neq i)} \frac{\tilde{U}_k - \tilde{U}_i}{\varepsilon_k - \varepsilon_i} = 0.
\end{align}
The sum of the rescaled EBV is twice the number of pairs
\begin{align}
\sum_i \tilde{U}_i = 2M.
\end{align}
When $g=0$, the reduced BCS Hamiltonian has no interaction and the RG states are Slater determinants defined by which spatial orbitals are occupied. Remarkably, as $g$ is increased, the RG states evolve uniquely and continuously from the $g=0$ solution. It is therefore unambiguous to label RG states \emph{at any finite $g$} based on their fixed occupations at $g=0$. In particular, the ground state of the reduced BCS Hamiltonian is \emph{always} the state 11..10...00 which has the pairs in the lowest spatial orbitals at $g=0$. The highest excited state is \emph{always} the state 00...01...1 which has the pairs in the highest spatial orbitals at $g=0$. Other states will cross at different values of $g$.

\begin{figure}[ht!]
	\begin{subfigure}{\textwidth}
		\includegraphics[width=0.475\textwidth]{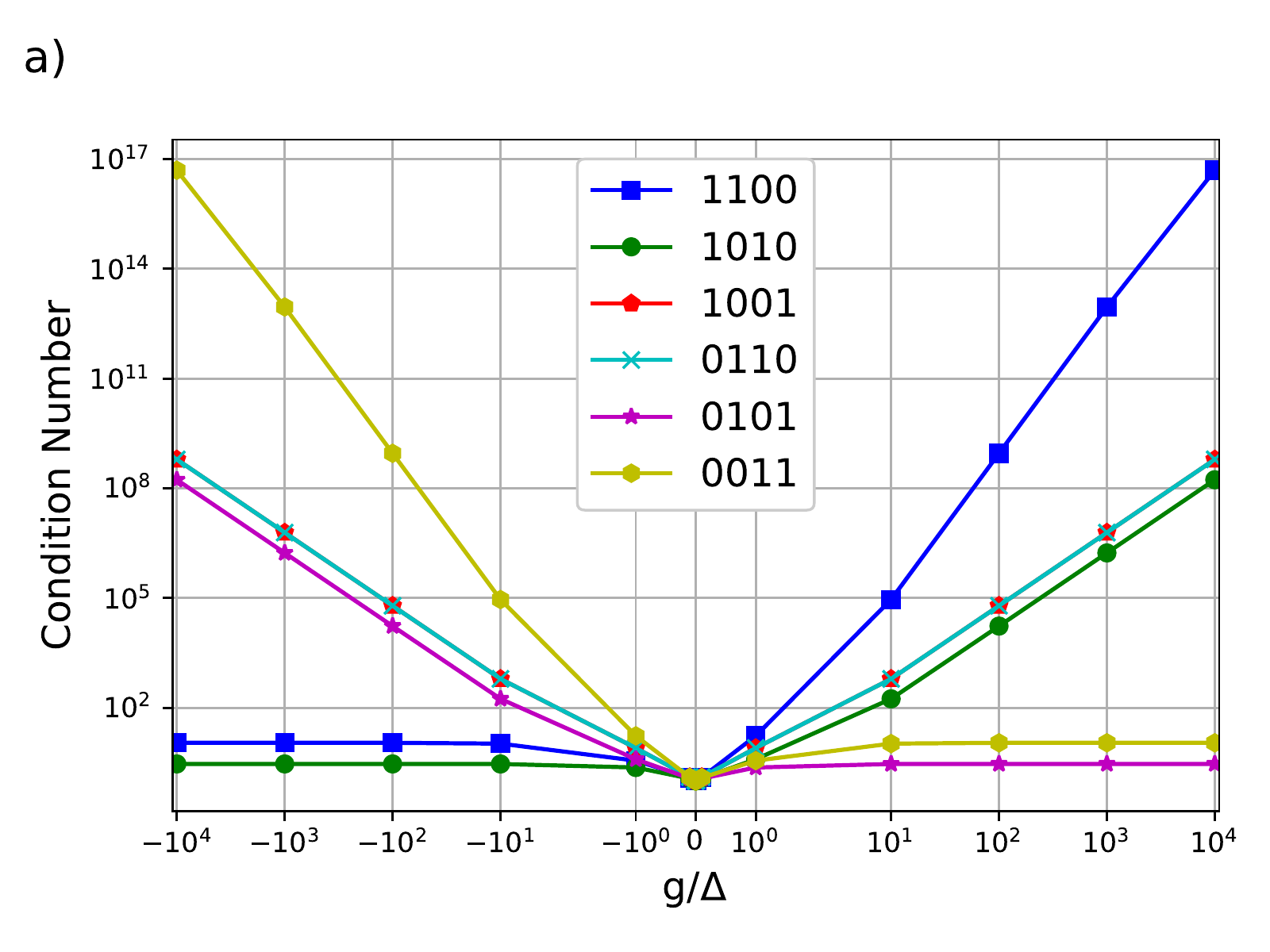} \hfill
		\includegraphics[width=0.475\textwidth]{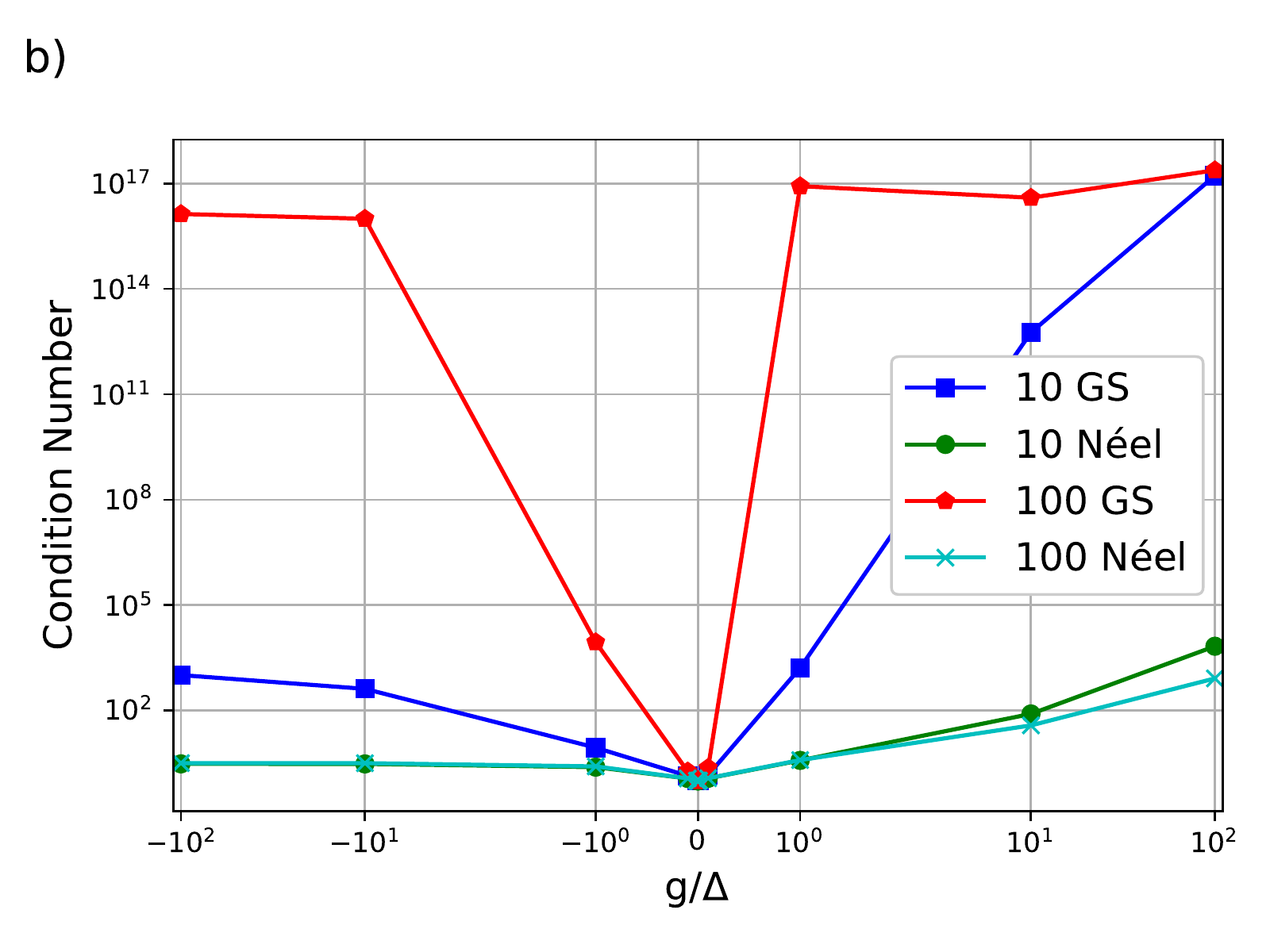}
	\end{subfigure}
		\caption{(a) Condition numbers for matrix $\bar{J}$ for each state in half-filled 4-site picket fence model as function of $g/\Delta$. (b) Condition numbers for matrix $\bar{J}$ for ground state and N\'{e}el state in half-filled 10- and 100-site picket fence model as function of $g/\Delta$.}
		\label{fig:PF_CN_curves}
\end{figure}

We will focus our efforts on two particular forms of reduced BCS Hamiltonians. First, the worst-case scenario is that the single-particle spectrum is completely degenerate. Thus we will employ the picket-fence (PF) model, a reduced BCS Hamiltonian with $\{\varepsilon\}$ evenly separated with spacing $\Delta$
\begin{align} \label{eq:pf_ham}
\hat{H}_{PF} =  \frac{1}{2} \sum^N_{i=1} (i-1) \Delta \hat{n}_i -\frac{g}{2} \sum_{ij} S^+_i S^-_j.
\end{align}
As $g/\Delta$ becomes very large, the single-particle energies effectively become completely degenerate, so we will compute the condition number as a function $g/\Delta$. Condition numbers are objects from numerical analysis that measure \emph{additional} loss of precision caused by ill-conditioning of the underlying matrix.\cite{trefethen_book_2} In this particular case we take the condition number of $\bar{J}$ to be the ratio of its largest to its smallest singular values. The condition number of $\bar{J}$ was computed for each RG state of a half-filled 4-site PF model and is plotted in figure \ref{fig:PF_CN_curves}a. From our variational calculations for bond-breaking processes,\cite{fecteau:2022} we have found that the optimal RG state is labelled 1010...10 which in the dissociation limit becomes the generalized-valence-bond (GVB) wavefunction
\begin{align}
\ket{\text{GVB}} = (S^+_1 - S^+_2)(S^+_3 - S^+_4)\dots (S^+_{2N-1} - S^+_{2N}) \ket{\theta}.
\end{align}
We refer to the 1010...10 RG state as the N\'{e}el RG state. Condition numbers for $\bar{J}$ were computed for half-filled 10- and 100-site PF models for the RG ground and N\'{e}el states, and are presented in figure \ref{fig:PF_CN_curves}b.

\begin{figure}[ht!]
	\begin{subfigure}{\textwidth}
		\includegraphics[width=0.475\textwidth]{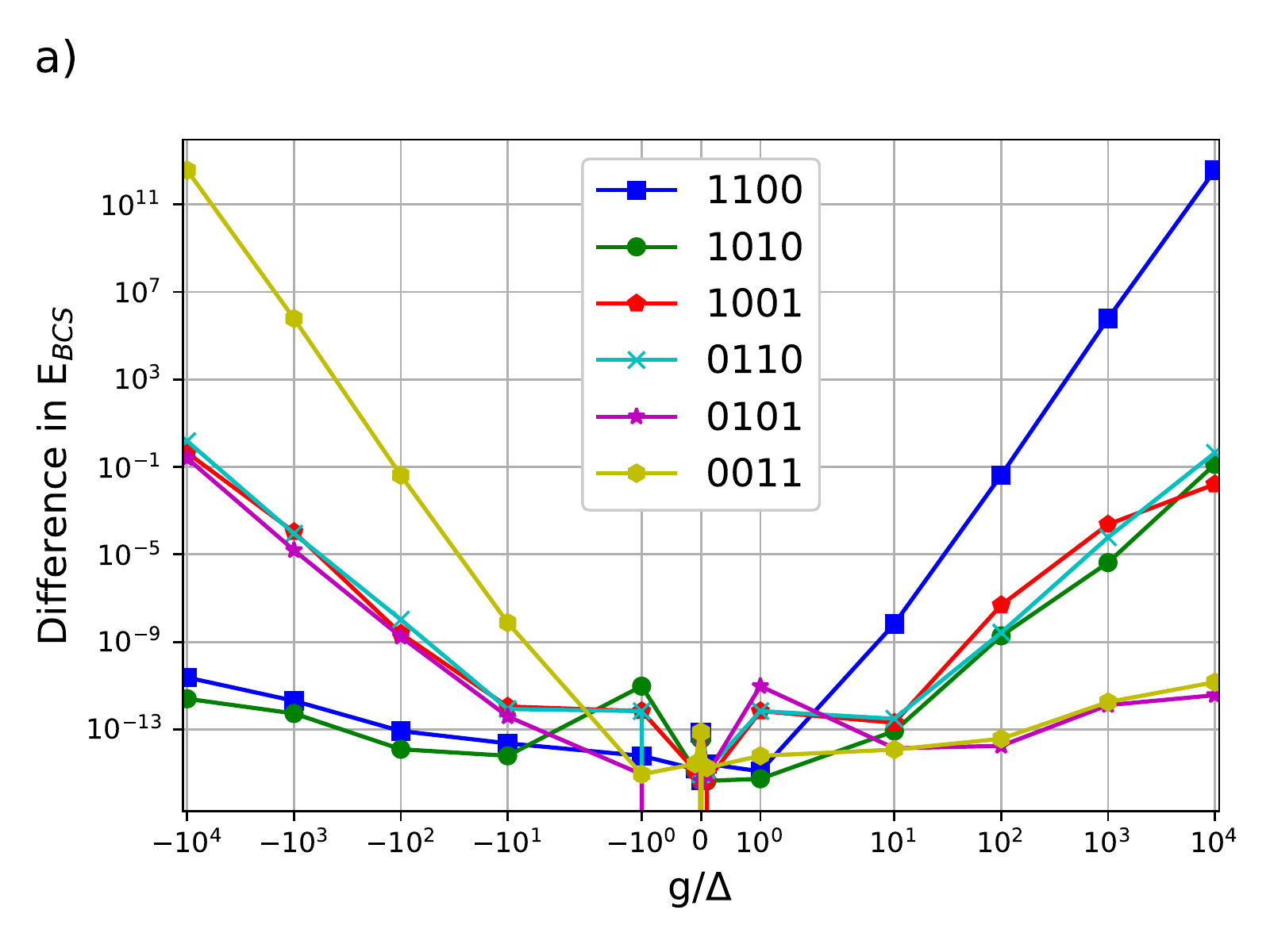} \hfill
		\includegraphics[width=0.475\textwidth]{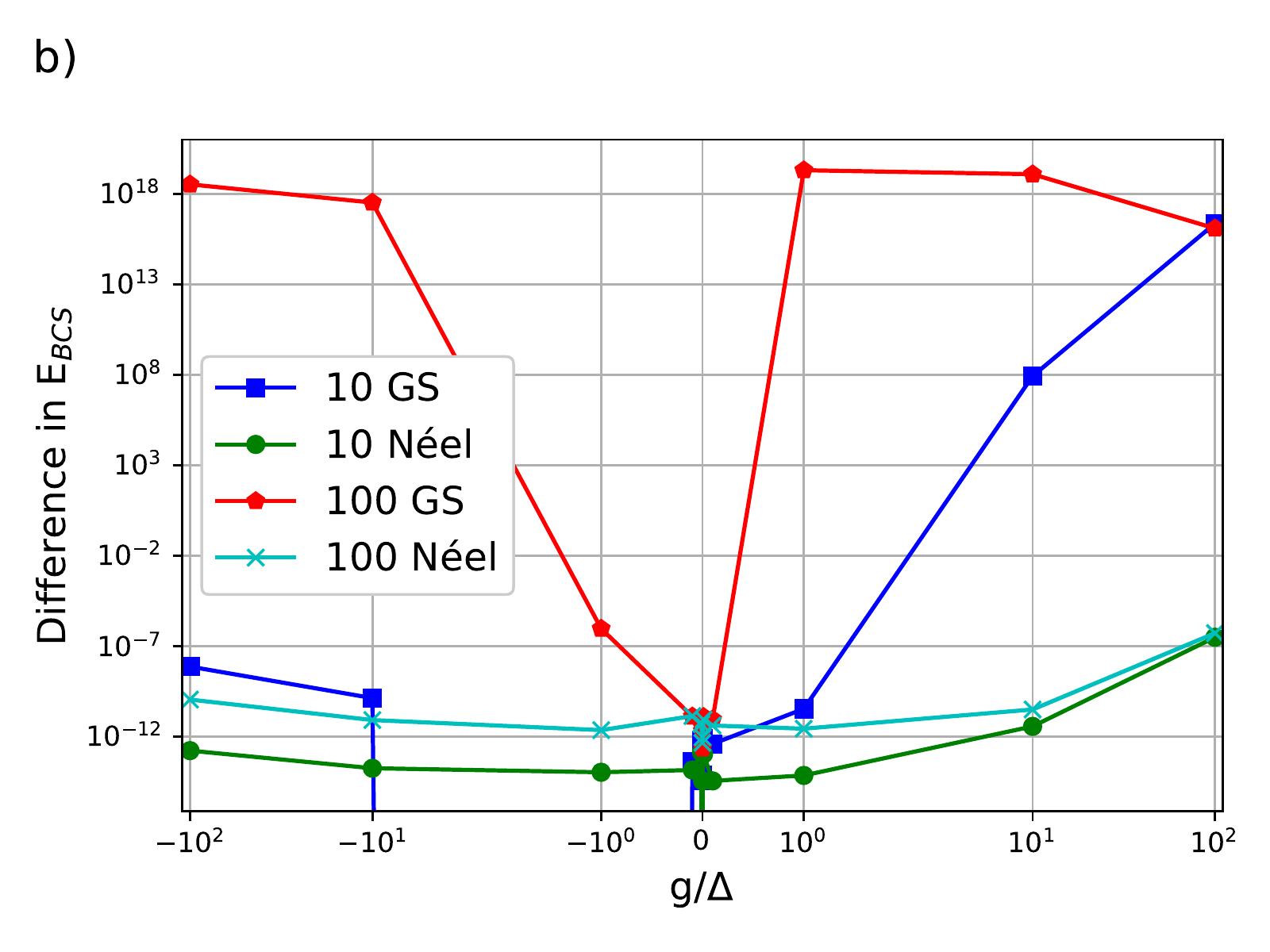}
	\end{subfigure}
		\caption{(a) Difference in $E_{BCS}$ energy expressions for each state in half-filled 4-site picket fence model as function of $g/\Delta$. (b) Difference in $E_{BCS}$ energy expressions for ground state and N\'{e}el state in half-filled 10- and 100-site picket fence model as function of $g/\Delta$.}
		\label{fig:PF_diff_curves}
\end{figure}

As can be seen in figure \ref{fig:PF_CN_curves}, the condition number grows with $g/\Delta$. How large a condition number is acceptable? This can be judged by the associated loss of precision in the computation of the RDM elements. In our previous papers, it was necessary to include a consistency check for the rapidity-based expressions. The elements $\gamma_k$ and $D_{kl}$ both satisfy trace conditions,
\begin{align}
\sum_k \gamma_k &= M \\
\sum_{kl} D_{kl} &= M(M-1)
\end{align}
where because we have set $D_{kk} = 0$, its trace is \emph{not} $M^2$. However, $P_{kl}$ does not have a simple criterion. To judge the overall loss of precision we computed the energy of the reduced BCS Hamiltonian \eqref{eq:bcs_ham} in two different ways. First, it is the sum of the rapidities
\begin{align} \label{eq:bcs_en}
E_{BCS} = \sum^M_{a=1} u_M
\end{align}
but it can also be computed as
\begin{align} \label{eq:rap_check}
E_{BCS} = \sum^N_{k=1} \varepsilon_k \gamma_k - \frac{g}{2}\sum_{kl} P_{kl}.
\end{align}
In the course of our variational optimizations, if the expressions \eqref{eq:bcs_en} and \eqref{eq:rap_check} differed by more than $1\times 10^{-6}$ Hartree, the loss of precision was deemed to be too large and the point was rejected. The purpose of the present contribution is to employ the EBV without the rapidities. The energy can also be computed in terms of the EBV as
\begin{align} \label{eq:ebv_check}
E_{BCS} = \frac{g}{2}M(M-N-1) + \frac{1}{2} \sum^N_{i=1} \varepsilon_i \tilde{U}_i.
\end{align}
The disagreement between \eqref{eq:rap_check} and \eqref{eq:ebv_check} was computed for \eqref{eq:pf_ham} as a function of $g/\Delta$, and is shown in figure \ref{fig:PF_diff_curves}.
The increase in condition number and the disagreement in $E_{BCS}$ are correlated. A condition number on the order of $10^5$ appears to be the limit of what we judge to be acceptable, i.e. if the condition number of $\bar{J}$ is larger than $10^5$, the levels are effectively degenerate and must be treated as such directly. Again, completely degenerate PF models are the worst-case scenario. 

\begin{figure}[ht!]
	\begin{subfigure}{\textwidth}
		\includegraphics[width=0.475\textwidth]{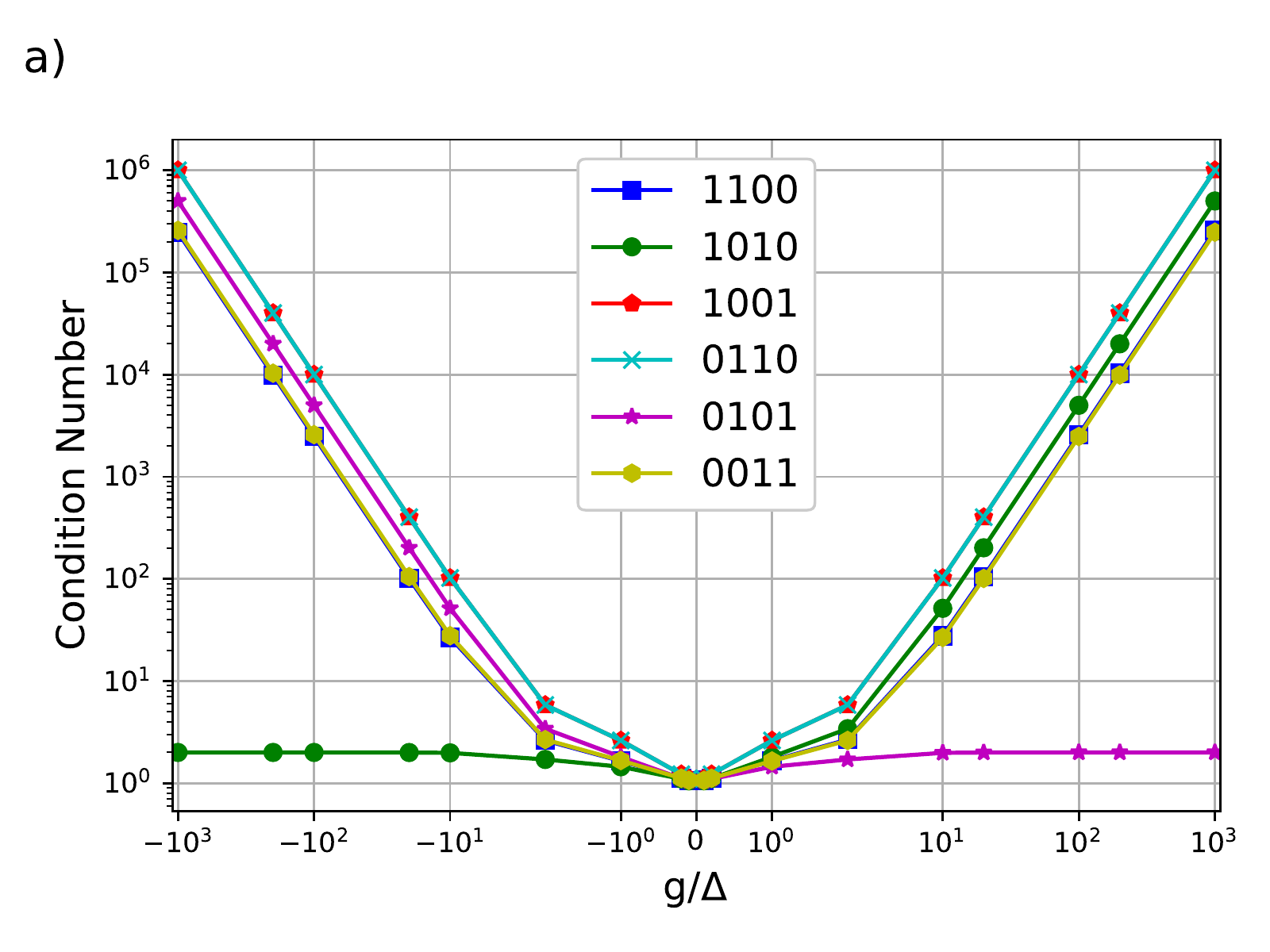} \hfill
		\includegraphics[width=0.475\textwidth]{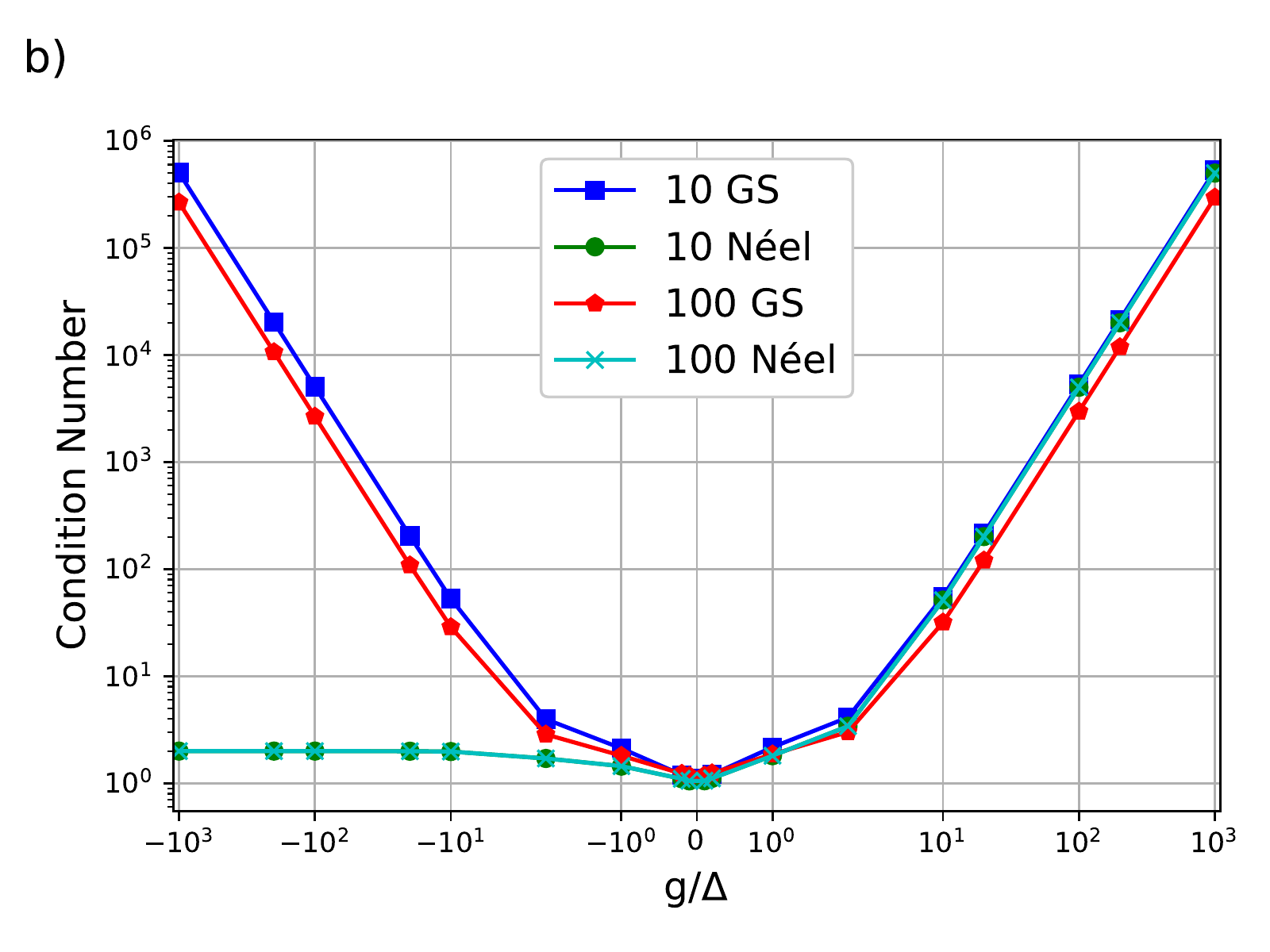}
	\end{subfigure}
	\begin{subfigure}{\textwidth}
		\includegraphics[width=0.475\textwidth]{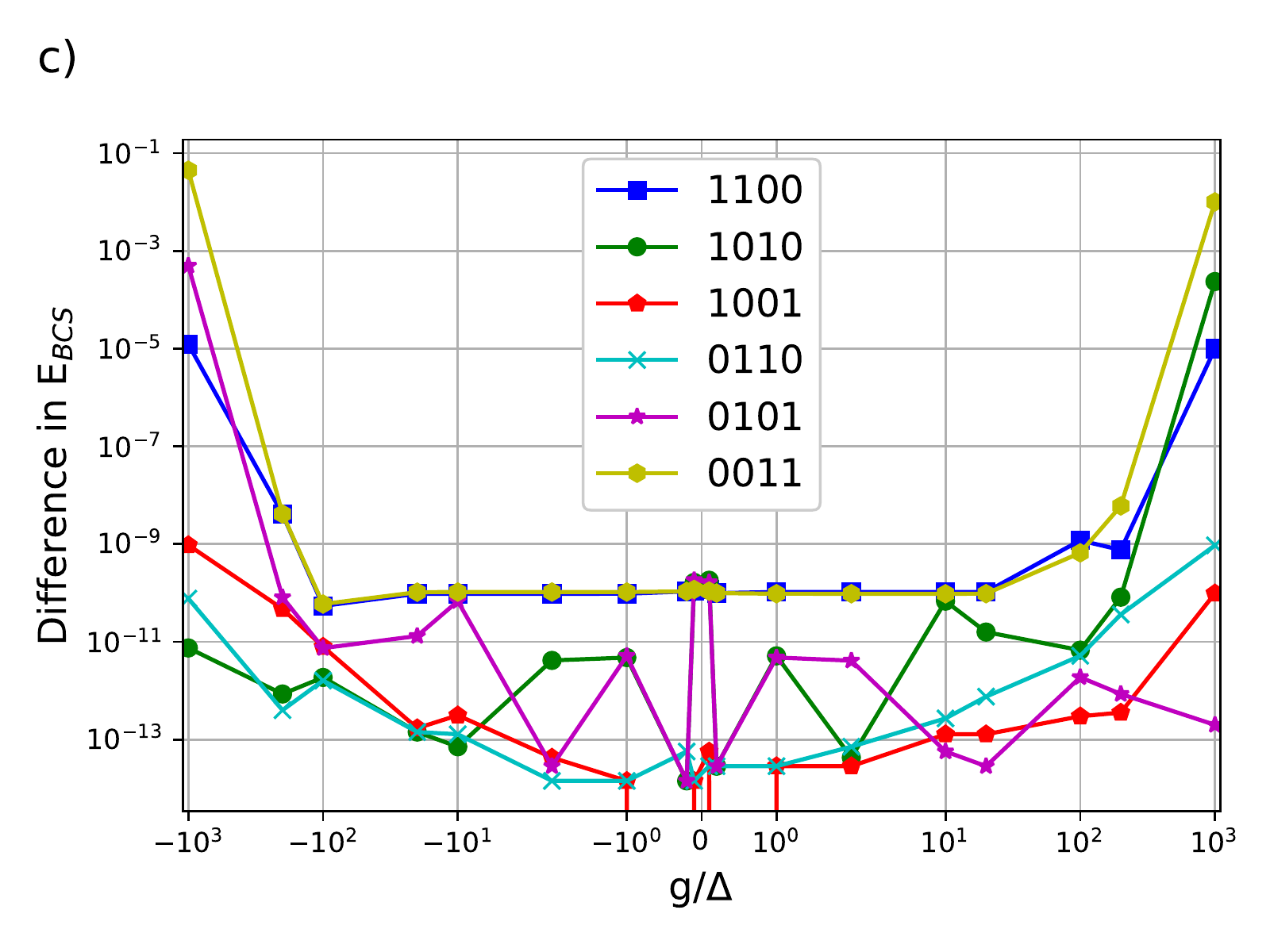} \hfill
		\includegraphics[width=0.475\textwidth]{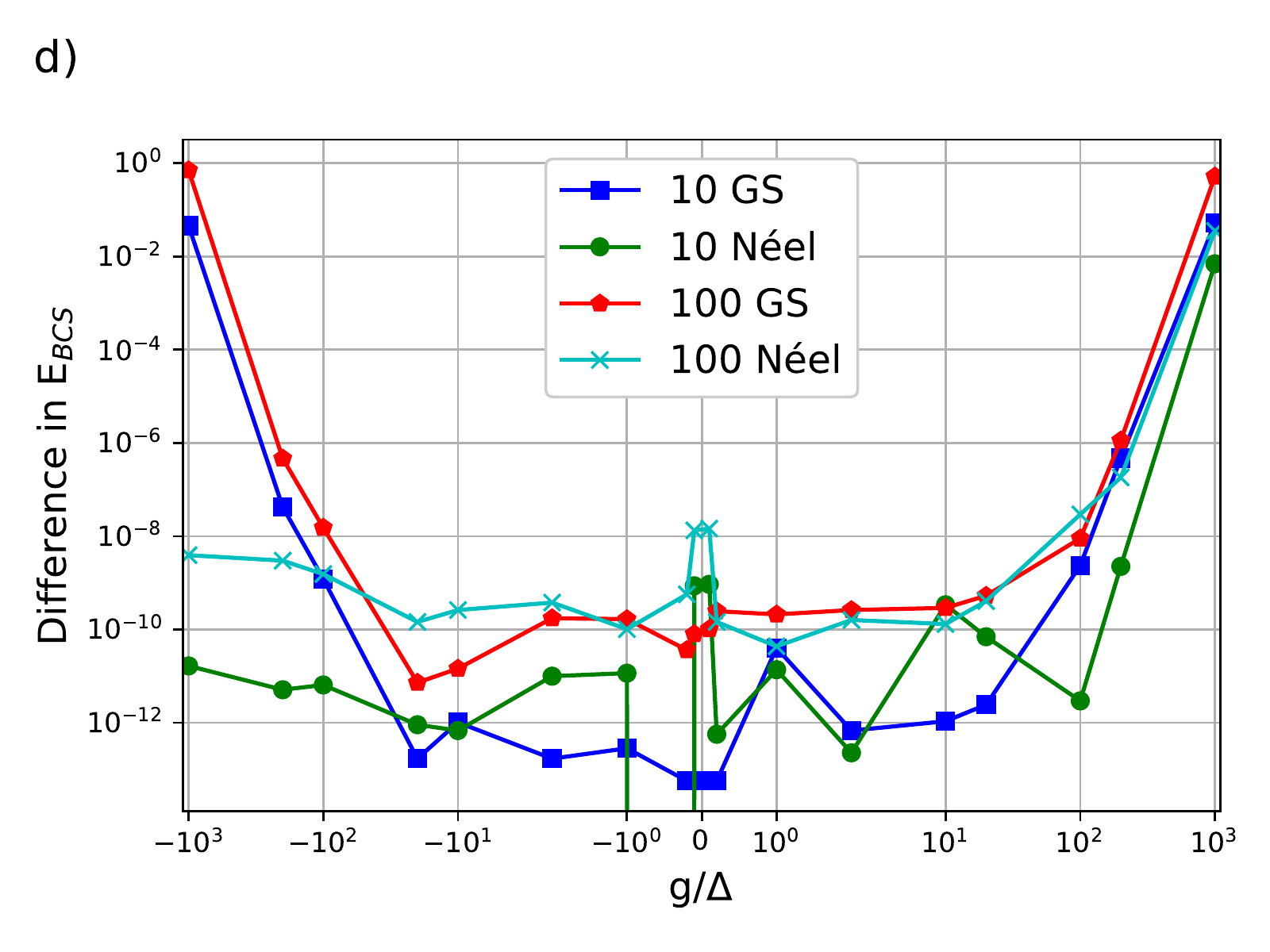}
	\end{subfigure}
		\caption{(a) Condition numbers for matrix $\bar{J}$ for each state in half-filled 4-site VB type model as function of $g/\Delta$. (b) Condition numbers for matrix $\bar{J}$ for ground state and N\'{e}el state in half-filled 10- and 100-site VB type model as function of $g/\Delta$. (c) Difference in $E_{BCS}$ energy expressions for each state in half-filled 4-site VB type model as function of $g/\Delta$. (b) Difference in $E_{BCS}$ energy expressions for ground state and N\'{e}el state in half-filled 10- and 100-site VB type model as function of $g/\Delta$.}
		\label{fig:VB_curves}
\end{figure}

The second particular type of reduced BCS Hamiltonian we consider is motivated by the N\'{e}el RG states we found for molecular dissociations. For such states, the reduced BCS Hamiltonian parameters we found were pairs of near-degenerate $\{\varepsilon\}$ well-separated in energy. Thus, we consider a valence-bond (VB) type reduced BCS Hamiltonian
\begin{align}
\hat{H}_{VB} = \frac{1}{2} \sum^N_{i=1} \left( (\varepsilon_i - \Delta) \hat{n}_{2i} + (\varepsilon_i + \Delta )\hat{n}_{2i+1} \right) -\frac{g}{2} \sum_{ij} S^+_i S^-_j
\end{align}
and again compute the condition number of $\bar{J}$ as a function of the ratio $g/\Delta$. In particular, as there is a potential for ambiguity, the pairing strength $g$ is held fixed at $\pm 1$ while the spacing $\Delta$ is varied. The single-particle energies have a constant spacing of 100$g$. Condition numbers as well as the difference in the expressions \eqref{eq:rap_check} and \eqref{eq:ebv_check} were computed for half-filled 4-, 10- and 100-site VB type models and are plotted in figure \ref{fig:VB_curves}.
Again, a condition number on the order of $10^5$ appears to cause an unacceptable loss of precision. The states that interest us most are the N\'{e}el states, in particular for repulsive (negative) interactions. Remarkably, even for very large values of $g/\Delta$, the corresponding condition number remains small. This means that it is acceptable to naively treat the levels as being non-degenerate even if their splitting is very small. However, in such cases solving the EBV equations \eqref{eq:ebv_eq} becomes much more expensive.\cite{fecteau:2022}

\section{Conclusion}
Simple and clean expressions for the RDM elements of RG states have been obtained in terms of the EBV. Solving for the rapidities is entirely avoided, reducing the computational cost and preventing loss of numerical precision. Unless the the number of pairs is very small compared with the number of spatial orbitals, there is no reason to employ rapidities at all. Analytic gradients and Hessians of the RG Coulomb energy functional are computable from derivatives of the RDM elements, which require only linear equations and matrix multiplication with the inverse-derivative formula. The matrix $\bar{J}$ has been shown to be well-conditioned except in the limit of degeneracies in the single-particle energies. In this case, the approach to solve for the EBV must be modified substantially. The EBV scalar products may still be possible, but the generalization is rather complicated and will therefore be addressed in a further contribution.

Rapidity-based RDM expressions reduced to ratios of determinants differing by 2 columns, which further reduced to $2 \times 2$ determinants of single-column replacements. These are obtainable as solutions of linear equations, with the Gaudin matrix, through Cramer's rule. For EBV, the RDM elements reduce to summations over scaled second cofactors of the matrix $\bar{J}$, which, again, reduce to $2 \times 2$ determinants of scaled first cofactors through Jacobi's theorem. Rather than compute cofactors directly, we noticed that through the adjugate formula the scaled first cofactors of $\bar{J}$ are exactly the elements of $\bar{J}^{-1}$. Everything required comes from $\bar{J}^{-1}$ and systems of linear equations with $\bar{J}$. 

\section{Acknowledgements}
P.A.J. was supported by NSERC and Compute Canada. We gratefully thank the Laboratoire de Physique et Chimie Th\'{e}oriques at the Universit\'{e} de Lorraine for additional support as a visiting professor.

\section{Data Availability}
The data that support the findings of this study are available from the corresponding author upon reasonable request.

\appendix

\section{Determinant Identities} 
\subsection{Proof of lemma \ref{lem:mat_diag}} \label{sec:lemma_1}
The proof is straightforward and is a direct consequence of the fact that for arbitrary $a$, $b$, $z$
\begin{align} \label{eq:frac_prop}
\frac{1}{a-b} \frac{b-z}{a-z} = \frac{1}{a-b} - \frac{1}{a-z}.
\end{align}
First 
\begin{align}
\det (J - d(z)) &= \det \begin{pmatrix}
J_{11} - \frac{1}{\varepsilon_1 - z} & \frac{1}{\varepsilon_1 - \varepsilon_2} & \dots & \frac{1}{\varepsilon_1 - \varepsilon_N} \\
\frac{1}{\varepsilon_2 - \varepsilon_1} & J_{22} - \frac{1}{\varepsilon_2 - z} & \dots & \frac{1}{\varepsilon_2 - \varepsilon_N} \\
\vdots & \vdots & \ddots & \vdots \\
\frac{1}{\varepsilon_N - \varepsilon_1} & \frac{1}{\varepsilon_2 - \varepsilon_N} & \dots & J_{NN} - \frac{1}{\varepsilon_N - z} 
\end{pmatrix},
\end{align}
now each row $i$ is scaled by $\frac{1}{\varepsilon_i - z}$ while each column $j$ is scaled by $\varepsilon_j -z$, effectively multiplying the expression by 1, giving
\begin{align}
\det (J - d(z)) &= \det \begin{pmatrix}
J_{11} - \frac{1}{\varepsilon_1 - z} & \frac{1}{\varepsilon_1 - \varepsilon_2} \frac{\varepsilon_2-z}{\varepsilon_1 - z} & \dots & \frac{1}{\varepsilon_1 - \varepsilon_N} \frac{\varepsilon_N-z}{\varepsilon_1 - z} \\
\frac{1}{\varepsilon_2 - \varepsilon_1}\frac{\varepsilon_1-z}{\varepsilon_2 - z} & J_{22} - \frac{1}{\varepsilon_2 - z} & \dots & \frac{1}{\varepsilon_2 - \varepsilon_N} \frac{\varepsilon_N-z}{\varepsilon_2 - z} \\
\vdots & \vdots & \ddots & \vdots \\
\frac{1}{\varepsilon_N - \varepsilon_1}\frac{\varepsilon_1-z}{\varepsilon_N - z} & \frac{1}{\varepsilon_N - \varepsilon_2}\frac{\varepsilon_2-z}{\varepsilon_N - z} & \dots & J_{NN} - \frac{1}{\varepsilon_N - z} 
\end{pmatrix}.
\end{align}
Using \eqref{eq:frac_prop} gives
\begin{align}
\det (J - d(z)) &= \det \begin{pmatrix}
J_{11} - \frac{1}{\varepsilon_1 - z} & \frac{1}{\varepsilon_1 - \varepsilon_2} - \frac{1}{\varepsilon_1 - z} & \dots & \frac{1}{\varepsilon_1 - \varepsilon_N} - \frac{1}{\varepsilon_1 - z} \\
\frac{1}{\varepsilon_2 - \varepsilon_1} - \frac{1}{\varepsilon_2 - z} & J_{22} - \frac{1}{\varepsilon_2 - z} & \dots & \frac{1}{\varepsilon_2 - \varepsilon_N} - \frac{1}{\varepsilon_2 - z} \\
\vdots & \vdots & \ddots & \vdots \\
\frac{1}{\varepsilon_N - \varepsilon_1} - \frac{1}{\varepsilon_N - z} & \frac{1}{\varepsilon_2 - \varepsilon_N} - \frac{1}{\varepsilon_N - z} & \dots & J_{NN} - \frac{1}{\varepsilon_N - z} 
\end{pmatrix} \\
&= \det (J - \mathbf{x}(z) \mathbf{1}^T)
\end{align}
as quoted in lemma \ref{lem:mat_diag}.

\subsection{Proof of lemma \ref{lem:update}} \label{sec:lemma_2}
The proof follows from the matrix determinant lemma for rank-one updates followed by cleaning up with column operations. In particular, the matrix determinant lemma states, for an invertible $J$, the result of a rank-one update is
\begin{align}
\det (J - \textbf{x}\textbf{y}^T) = \det J \left( 1 - \textbf{y}^T J^{-1} \textbf{x} \right).
\end{align}
The weighted sum of rank-one updates is then
\begin{align}
\sum^M_{j=1} \alpha_j \det ( J - \textbf{x}_j \textbf{y}^T) &= \det(J) \left( \sum^M_{j=1} \alpha_j \right)  - \det(J) \left( \textbf{y}^T J^{-1}\tilde{\textbf{x}} \right)
\end{align}
with $\tilde{\textbf{x}} \equiv \sum^M_{j=1} \alpha_j \textbf{x}_j$. With the matrix determinant lemma again, 
\begin{align}
\det (J - \tilde{\textbf{x}} \textbf{y}^T) &=  \det J \left( 1 -   \textbf{y}^T J^{-1}\tilde{\textbf{x}} \right),
\end{align}
so that
\begin{align}
\sum^M_{j=1} \alpha_j \det ( J - \textbf{x}_j \textbf{y}^T) = \det(J) \left( \sum^M_{j=1} \alpha_j - 1 \right) + \det (J - \tilde{\textbf{x}} \textbf{y}^T).
\end{align}
Both of these $N \times N$ determinants are lifted to $N+1 \times N+1$ by adding a first row with one non-zero entry. This introduces an \emph{arbitrarily} chosen first column which we choose to be zeros for the first determinant and $\tilde{\textbf{x}}$ in the second, giving
\begin{align}
\sum^M_{j=1} \alpha_j \det ( J - \textbf{x}_j \textbf{y}^T) =
\left( \sum^M_{j=1} \alpha_j - 1 \right)
\det \left(
\begin{array}{ccccc}
1 & 0 & \dots & 0 \\
0  & \\
\vdots & & J &  \\
0 
\end{array} \right)
+ \det \left(
\begin{array}{ccccc}
1 & 0 & \dots & 0 \\
\tilde{x}_1  & \\
\vdots & & J -\tilde{\textbf{x}}\textbf{y}^T &  \\
\tilde{x}_N 
\end{array} \right)
\end{align}
Now, $y_j$ times the first column is added to each other column
\begin{align}
\sum^M_{j=1} \alpha_j \det ( J - \textbf{x}_j \textbf{y}^T) =
\left( \sum^M_{j=1} \alpha_j - 1 \right)
\det \left(
\begin{array}{ccccc}
1 & y_1 & \dots & y_N \\
0  & \\
\vdots & & J &  \\
0 
\end{array} \right)
+ \det \left(
\begin{array}{ccccc}
1 & y_1 & \dots & y_N \\
\tilde{x}_1  & \\
\vdots & & J  &  \\
\tilde{x}_N 
\end{array} \right)
\end{align}
and we arrive at two determinants which differ in one column, which can be condensed to one
\begin{align}
\sum^M_{j=1} \alpha_j \det ( J - \textbf{x}_j \textbf{y}^T) &=
\det \left(
\begin{array}{ccccc}
\sum^M_{j=1} \alpha_j & y_1 & \dots & y_N \\
\tilde{x}_1  & \\
\vdots & & J  &  \\
\tilde{x}_N 
\end{array} \right) \\
&= \det \begin{pmatrix}
\sum^M_{j=1} \alpha_j & \textbf{y}^T \\
\tilde{\textbf{x}} & J
\end{pmatrix},
\end{align}
which is the desired result. 

\bibliography{xxx_ebv_arXiv}

\bibliographystyle{unsrt}

\end{document}